\documentclass[]{pasj02} 
\usepackage[switch,mathlines]{lineno} 
\usepackage{graphicx}
\usepackage{dcolumn}
\usepackage{bm}
\usepackage[utf8]{inputenc}
\usepackage[T1]{fontenc}
\usepackage{mathptmx}
\usepackage{etoolbox}
\usepackage{subcaption}
\usepackage{color}
\usepackage[whole]{bxcjkjatype}
\usepackage{siunitx}
\usepackage{comment}
\usepackage{url}
\usepackage{soul}
\usepackage{physics}
\usepackage{bm}
\usepackage[normalem]{ulem}
\usepackage{xcolor}

\jyear{2025}
\Received{}
\Accepted{}

\begin{document} 

\title{Influence of kinetic effects in large-scale magnetic reconnection with multi-hierarchy simulation code KAMMUY}

\author{
 Keita \textsc{Akutagawa},\altaffilmark{1}\altemailmark\orcid{0009-0005-4643-1660} \email{akutagawa@eps.s.u-tokyo.ac.jp} 
 Shinsuke \textsc{Imada},\altaffilmark{1}\orcid{0000-0001-7891-3916}
 and 
 Munehito \textsc{Shoda}\altaffilmark{1}\orcid{0000-0002-7136-8190}
}
\altaffiltext{1}{Department of Earth and Planetary Science, Graduate School of Science, The University of Tokyo, 7-3-1 Hongo, Bunkyo-ku, Tokyo 113-0033, Japan}


\KeyWords{methods: numerical, magnetic reconnection, plasmas, Sun: flares}  

\maketitle

\begin{abstract}
Magnetic reconnection is a multiscale phenomenon where fluid- and particle-scale processes interact. The particle-in-cell (PIC) method, capable of resolving kinetic (particle-scale) physics, is extensively used to study the kinetic effects in magnetic reconnection. Meanwhile, because of the high computational cost, PIC simulations cannot capture the interaction between kinetic and fluid dynamics, which poses a major obstacle to understanding magnetic reconnection in large-scale phenomena such as solar flares. A multi-hierarchy simulation that combines Magnetohydrodynamics (MHD) and PIC provides a promising means to overcome these spatial and temporal scale gaps. We developed a multi-hierarchy simulation code KAMMUY (Kinetic And Magnetohydrodynamic MUlti-hierarchY simulation code), in which an ideal MHD simulation for a large domain and a PIC simulation for a smaller domain are solved in parallel with mutual information exchange. To validate the code, we conducted test simulations of MHD wave propagation and the shock tube problem. The results demonstrate that short-wavelength, high-frequency waves generated in the PIC region do not propagate into the MHD region, whereas MHD-scale structures propagate smoothly into the PIC region, highlighting the capability of our code for numerical studies of magnetic reconnection. By applying the KAMMUY code to magnetic reconnection while varying the PIC domain size, we find that the reconnection rate remains unchanged, regardless of the extent of the PIC region where the Hall magnetic field is present. It suggests that the spatial extension of the Hall magnetic field on the scale of $10 \sim 100 \lambda_i$ does not influence the reconnection rate.
\end{abstract}



\section{Introduction}\label{chap1:introduction}


Plasma phenomena often show multiscale behavior due to the interaction between fluid- and particle-scale physics \citep{squire2022}. Magnetohydrodynamics (MHD) describes the fluid scale physics, while kinetic theory represents the particle scale physics, and their self-consistent treatment is crucial for understanding multiscale plasma dynamics. In simulations, particle-in-cell (PIC) methods are commonly used for kinetic plasmas. Integration of MHD and PIC models within one framework is promising but unresolved challenge. Some methods have been developed to date, however there is no established computational approach for this coupling.

One of the main motivations for bridging the gap between fluid and particle scales is to clarify the multiscale nature of magnetic reconnection. Magnetic reconnection is a fluid-scale phenomenon observed in astrophysical and laboratory plasmas, while its timescale and reconnection rate are controlled by resistivity arising from plasma kinetics. Several models have been proposed under the resistive MHD framework, including the Petschek reconnection model \citep{petschek1965, ugai1995, zenitani2011petschek, zenitani2015}, the plasmoid-mediated reconnection model \citep{shibata2001, loureiro2007, bhattacharjee2009, huang2011, zenitani2020}, and the dynamical Petschek reconnection model \citep{shibayama2015, shibayama2019}. Although these models can partly explain the observed reconnection rate in solar flares, the resistivity is phenomenologically prescribed, and it is difficult to determine which model is most plausible. This limitation poses a major obstacle in reconnection studies of fluid-scale astrophysical systems. To overcome this, MHD simulations incorporating resistivity derived self-consistently from plasma kinetics are required.

It is well established that the Hall effect is essential to explain the observed reconnection rate of $\mathcal{O}(0.1)$ \citep{birn2001}. Most simulation studies, however, have been limited to simulation box sizes of $\mathcal{O}(100 \lambda_i)$, where $\lambda_i$ denotes the ion inertial length. In such systems, Hall magnetic fields with quadrupole structure persist and span the entire domain without damping \citep{fujimoto2006, drake2009, zenitani2011pop, walia2022}. These results suggest that the Hall effect plays a key role in systems of $\mathcal{O}(100\lambda_i)$ or larger. Nevertheless, its overall impact on the efficiency of the energy conversion in entire systems (=: global reconnection rate), which is not the reconnection rate achieved locally, remains unclear. It is uncertain whether the Hall effect is only important near the localized diffusion region or also significant at global scales. In particular, in the astrophysical context, whether the kinetic and two-fluid (including Hall) effects control the reconnection rate in solar flares of $\mathcal{O}(10^{7-8}\lambda_i)$ remains an open and critical question \citep{shay2024arXiv, nakamura2025, drake2025}. To properly evaluate the role of the Hall effect in the global reconnection rate, it is therefore necessary to use models capable of describing full plasma kinetic effects in fluid-scale systems.

In order to simulate as large systems as possible with the first-principle plasma simulation method PIC, several approaches have been developed. The fully implicit PIC code iPIC3D proposed by \citet{markidis2010} relaxes the time-step restriction by employing implicit integration for particles and fields. This enables simulations in regimes inaccessible to explicit PIC, such as the realistic mass ratio $m_i/m_e \sim 1836$ and the large frequency ratio $\omega_{pe}/\Omega_{ce} \gg 1$. Nevertheless, resolving electron scales in space and time still incurs a computational cost comparable to explicit PIC. The Energy Conserving semi-implicit method (ECsim) proposed by \citet{lapenta2017} addresses the energy dissipation problem in fully implicit PIC. By updating particles explicitly while solving fields implicitly, ECsim reduces the cost of the particle solver. It has been implemented in the PIC module of a multi-hierarchy framework \citep{zhou2019}. Another approach is AMR-PIC \citep{fujimoto2006}, which dynamically adapts grid resolution using particle splitting. This method enabled long-term simulations of magnetic reconnection in a $\mathcal{O}(650) \lambda_i$ 2D system \citep{fujimoto2018}. The various approaches shown above enables about 10 times larger scale systems compared to explicit PIC.

In this study, we focus on multi-hierarchy simulations. Originally introduced by \citet{sugiyama2007}, this approach solves MHD and PIC simultaneously, providing a promising means to bridge the large spatial and temporal gaps between fluid and particle scales. By applying PIC only in regions where kinetic effects are essential and using MHD elsewhere, the method reduces computational cost while preserving physical fidelity. Several implementations have since been proposed and validated: the original coupling of 1D explicit PIC with ideal MHD using a variable grid size ratio \citep{sugiyama2007}; a coupling of 2D explicit PIC with ideal MHD on the same grid \citep{usami2013, ogawa2016}; 3D full implicit PIC or ECsim coupled with Hall or ideal MHD \citep{daldorff2014, toth2017, dion2023, wang2024, makwana2017}; and a recent model combining semi-implicit PIC (implicit only for Maxwell’s equations) with ideal MHD \citep{haahr2025}. Some previous studies performed global earth magnetospheric simulation \citep{daldorff2014, toth2017, wang2024} embedding PIC in magnetopause and magnetotail reconnection regions. Multi-hierarchy simulation is still developing and promising approach to understand multiscale plasma dynamics.

Comparing to PIC method, multi-hierarchy method has the potential to reduce a lot of particles. Reducing particles is essential for computational memory and time because the most time-consuming parts of PIC are the calculation of particle moment (or current) and the MPI communication of particles. The ratio of the number of particles which can be reduced by multi-hierarchy scheme relative PIC in 2D system can be estimated as follows:
\begin{align}
    \frac{N_{\text{PIC}}}{N_{\text{tot}}} \sim \frac{N_{x, \text{PIC}} \times N_{y, \text{PIC}}}{N_{x, \text{MHD}} \times N_{y, \text{MHD}}} \times \left( \frac{\Delta_{\text{PIC}}}{\Delta_{\text{MHD}}} \right)^2,
\end{align}
where $N_{\text{PIC}}$ is the number of particles in PIC region of multi-hierarchy simulation ($\propto N_{x, \text{PIC}} \times N_{y, \text{PIC}}$) and $N_{\text{tot}}$ is the number of particles needed in entire region ($\propto N_{x,\text{tot}} \times N_{y,\text{tot}}$). $\Delta_{\text{MHD}}$ and $\Delta_{\text{PIC}}$ are the grid sizes of MHD and PIC. $N_{x,\text{tot}}$ and $N_{y,\text{tot}}$ are the number of MHD grid converted to PIC grid; $N_{x,\text{tot}} = N_{x, \text{MHD}} \times \Delta_{\text{MHD}} / \Delta_{\text{PIC}}$ and $N_{y,\text{tot}} = N_{y, \text{MHD}} \times \Delta_{\text{MHD}} / \Delta_{\text{PIC}}$. This estimation is strictly satisfied when particles are uniformly distributed. Using larger grid size ratio $\Delta_{\text{MHD}} \gg \Delta_{\text{PIC}}$ is essential for reducing computational cost.

We have developed a multi-hierarchy simulation code from scratch which connects 2D explicit PIC and ideal MHD with a variable grid size ratio, and named it KAMMUY (Kinetic And Magnetohydrodynamic MUlti-hierarchY simulation code). By employing an explicit PIC, our approach allows accurate treatment of electron scale physics. However, the strong numerical noise inherent in explicit PIC requires careful handling when coupling to the MHD. In addition, special treatment is needed to address the problem of non-zero divergence of magnetic field ($\nabla \cdot \bm{B} \neq 0$) that can arise when sending magnetic field from PIC to MHD. The details of schemes employed in KAMMUY are described in section \ref{chap2:schemes_of_KAMMUY}. KAMMUY is implemented in CUDA C++ so that it supports GPUs parallelization. To validate its applicability to a variety of physical problems, we performed several test simulations, including energy conservation property, Alfv\'en wave propagation, fast mode wave propagation, and a shock tube problem. The results of these test simulations are presented in section \ref{chap3:test_simulations}. Building on this validation, we applied KAMMUY to magnetic reconnection in $100\lambda_i$ system where $\lambda_i$ is the ion inertial length, which is presented in section \ref{chap4:magnetic_reconnection}. By varying the size of the PIC region in the inflow direction, we effectively controlled the spatial extent of the region where kinetic or Hall effects work to investigate how these effects influence the reconnection rate.


\section{Numerical scheme of KAMMUY code}\label{chap2:schemes_of_KAMMUY}

Our multi-hierarchy scheme broadly follows the interlocking method of \cite{sugiyama2007}. However, there are some problems when extending their algorithm in 2D systems, such as $\div \bm{B}$ error which our numerical calculation becomes unstable. In this section we describe in detail what algorithms KAMMUY follows.

\subsection{Numerical scheme of ideal MHD simulation}\label{chap2:mhd}
The MHD equations solved by the KAMMUY code are the conventional ideal MHD equations, expressed as follows:
\begin{align}
    &\frac{\partial \rho}{\partial t} + \nabla \cdot (\rho \bm{v}) = 0, \\
    &\rho \frac{d\bm{v}}{dt} = -\nabla p + \bm{j} \times \bm{B}, \\
    &\bm{E} + \bm{v} \times \bm{B} = \bm{0}, \\
    &\bm{j} = \frac{\nabla \times \bm{B}}{\mu_0}, \\
     &\frac{\partial e}{\partial t} + \nabla \cdot ((e + p_T) \bm{v} - \bm{B}(\bm{v} \cdot \bm{B})) = 0,
    \label{chap2:mhd_eq}
\end{align}
where $\rho$ denotes the mass density, $\bm{v} = (u, v, w)$ the bulk velocity, $\bm{j} = (j_x, j_y, j_z)$ the current density, $\bm{B} = (B_x, B_y, B_z)$ the magnetic field, $\bm{E} = (E_x, E_y, E_z)$ the electric field, $p$ the pressure, $p_T := p + |\bm{B}|^2 / 2$ is the total pressure, and $e$ is the total energy density. $\mu_0$ is the magnetic permeability of vacuum.

The MHD equations are solved using the finite volume method with the HLLD Riemann solver \citep{miyoshi2005}, 2nd-order MUSCL reconstruction \citep{vanleer1979}, and a 2nd-order Runge-Kutta method. The solenoidal condition of the magnetic field is enforced using the projection method \citep{brackbill1980}. We have tested our MHD simulation code with 12 shock tube problems \citep{brio1988} and checked the discontinuity is solved without any spurious oscillation, with Orszag-Tang vortex \citep{orszag1979} and checked the solenoidal error is removed well.

\subsection{Numerical scheme of PIC simulation}\label{chap2:pic}

The governing equations of PIC simulation are the equations of motion and the Maxwell's equations. Equations of motion are expressed as follows: 
\begin{align}
    &\frac{d \bm{x}}{dt} = \bm{v}, \\
    &\frac{d}{dt} (\gamma m_s \bm{v}) = q_s \left( \bm{E} (\bm{x}) + \bm{v} \times \bm{B} (\bm{x}) \right),
\end{align}
and Maxwell's equations are expressed as follows: 
\begin{align}
    &\nabla \cdot \bm{E} = \frac{\rho_q}{\varepsilon_0}, \\
    &\nabla \cdot \bm{B} = 0, \\
    &\nabla \times \bm{E} = -\frac{\partial \bm{B}}{\partial t}, \\
    &\nabla \times \bm{B} = \mu_0 \bm{j} + \frac{1}{c^2} \frac{\partial \bm{E}}{\partial t}, 
\end{align}
where $\bm{x}$ and $\bm{v}$ are the position and velocity, and $\gamma$ is the Lorentz factor of the particle. $m_s$ and $q_s$ are the rest mass and charge of particle species $s$. $\bm{E} = (E_x, E_y, E_z)$ and $\bm{B} = (B_x, B_y, B_z)$ are the magnetic and electric field, $\rho_q$ and $\bm{j} = (j_x, j_y, j_z)$ are the charge density and current density. $\epsilon_0$ and $\mu_0$ are the dielectric constant and magnetic permeability in vacuum. 

The equations of motion and Maxwell's equations are connected through the following relationships:
\begin{align}
    \rho_q &= \sum_s q_s \int f_s d\bm{v}, \\
    \bm{j} &= \sum_s q_s \int f_s \bm{v} d\bm{v}, 
\end{align}
where the subscript $s$ denotes the species of the particles, and the $f_s$ denotes the distribution function of species $s$.

The equations of motion are solved using the relativistically extended Buneman–Boris method \citep{boris1970}, which ensures second-order accuracy in time and numerical stability except for ultra-relativistic particles. Maxwell's equations are solved with the Yee lattice \citep{yee1966} and the Leapfrog scheme, which are second-order accurate in space and time. Charge and current densities are obtained from the distribution function using linear interpolation (CIC) following \citet{birdsall_and_langdon_1991}.

Using the Yee lattice guarantees $\nabla \cdot \bm{B} = 0$ to the machine precision. Meanwhile, the Poisson noise arising from $\nabla \cdot \bm{E} \neq \rho_q / \epsilon_0$ must be eliminated at each step, for which we apply the Langdon–Marder correction \citep{marder1987, langdon1992}. This approach introduces a pseudo current that disperses and attenuates the noise. The modified time evolution of the electric field is given by
\begin{align}
\frac{\partial \bm{E}}{\partial t} = c^2 \nabla \times \bm{B} - \frac{\bm{j}}{\epsilon_0} + d \nabla F,
\end{align}
where
\begin{align}
F := \nabla \cdot \bm{E} - \rho_q.
\end{align}
The quantity $F$ then evolves according to the diffusion equation:
\begin{align}
\frac{\partial F}{\partial t} - d \nabla^2 F = -\left( \frac{\partial \rho_q}{\partial t} + \nabla \cdot \bm{j} \right).
\end{align}
Here, $d$ is a free parameter and determined according to simulation setup. When the time evolution is solved explicitly, the Langdon–Marder correction efficiently suppresses Poisson noise using only local data, making it highly suitable for parallel computation. We have tested our PIC simulation code with two stream instability and compared growth rate with linear theory, with Weibel instability and checked energy conversion ratio from kinetic to magnetic energy, and with collisionless magnetic reconnection and checked the reconnection rate and field structures (Figure 1, 2, and 3 in \cite{akutagawa2025}).

\subsection{Numerical scheme of MHD-PIC coupling}\label{chap2:multi-hierarchy_scheme}

Our study is motivated to understand multiscale nature of plasma dynamics, in particular magnetic reconnection. Multi-hierarchy method can capture interactions between plasma kinetics and MHD self-consistently. In this section we summarize the methodology for connecting MHD and PIC models.

\begin{figure}[!t]
    \includegraphics[width=\linewidth]{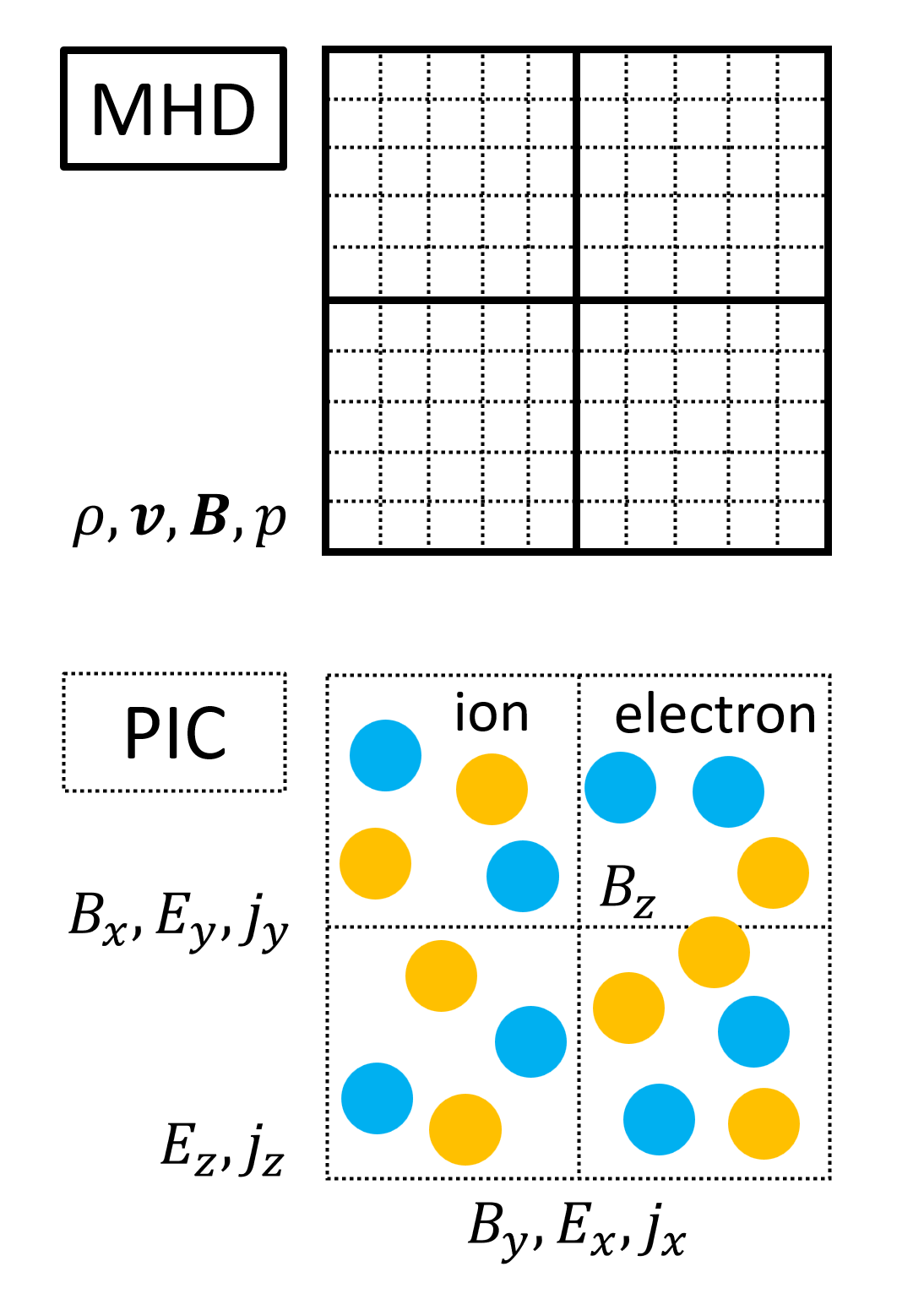}
    \caption{\label{chap2:fig:mhd_and_pic_grid}Schematic pictures of MHD and PIC grids. The upper panel shows the physical quantities of $\rho, \bm{v}, \bm{B}, p$ in MHD grid (solid line). The lower panel shows the physical quantities of $\bm{B}, \bm{E}, \bm{j}$ and particles (orange and light-blue circles) in PIC grid (dotted line). Dotted lines are drawn in MHD grid to emphasize that PIC grid is embedded in MHD region and is smaller than MHD grid.
    {Alt text: The $\rho, \bm{v}, \bm{B}$ and $p$ of MHD are at the cell center. The $E_z$ and $j_z$ of PIC are at the cell center. $B_y, E_x$ and $j_x$ of PIC are at the face center in $x$ direction. $B_x, E_y$ and $j_y$ of PIC are at the face center in $y$ direction. The $B_z$ of PIC is at the edge of cell.}}
\end{figure}

The locations of physical quantities in MHD and PIC are illustrated in Figure \ref{chap2:fig:mhd_and_pic_grid}. In MHD, all quantities are placed at the cell centre, whereas in PIC they are staggered. The multi-hierarchy scheme therefore requires a consistent and non-trivial mapping between the MHD and PIC quantities. In practice the PIC grid size $\Delta_{\text{PIC}}$ (comparable to the Debye length) must be much smaller than the MHD grid size $\Delta_{\text{MHD}}$. The freedom to choose $\Delta_{\text{MHD}} / \Delta_{\text{PIC}}$ is essential for numerical stability, as discussed in sub-subsection \ref{chap2:grid_size_and_time_step}.

\subsubsection{Exchange of physical quantities between MHD and PIC domains}\label{chap2:exchange_of_physical_quantities_between_mhd_and_pic}

\begin{figure}[!t]
    \includegraphics[width=\linewidth]{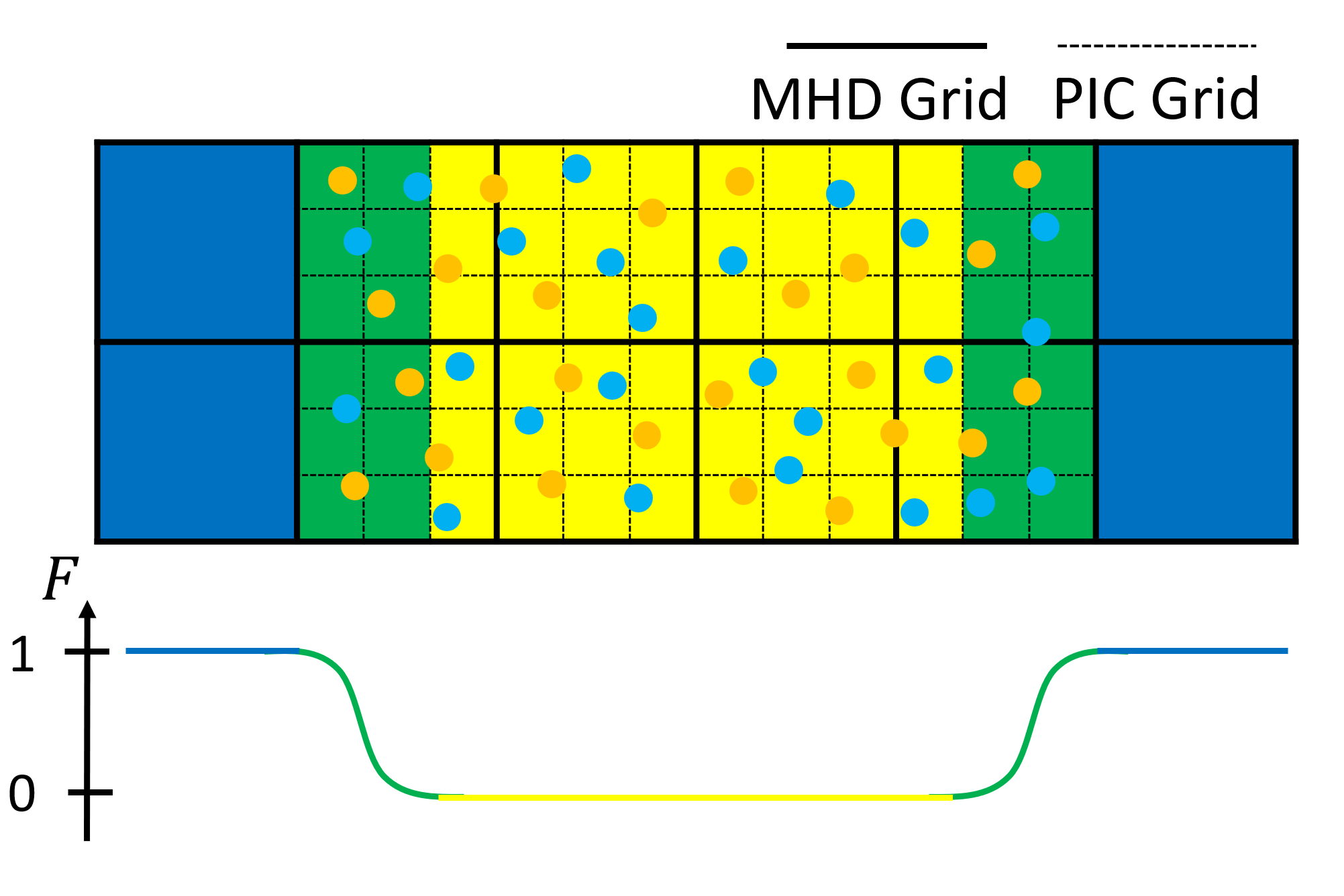}
    \caption{\label{chap2:fig:multi-hierarchy_scheme_space}Schematic picture of multi-hierarchy simulation in space. Solid and dotted line show MHD and PIC grids and orange and light-blue circles show particles in PIC. Blue area corresponds to the purely MHD region, yellow area corresponds to the purely PIC region, and green area is the interface region. PIC has the green and yellow region and MHD has the whole region.
    {Alt text: Schematic picture of multi-hierarchy simulation in space.}}
\end{figure}

To connect the MHD and PIC domains, we employ the interlocking method \citep{sugiyama2007, usami2013, daldorff2014, makwana2017} for exchanging physical quantities. In this approach, quantities within the boundary (interface) region are calculated as a weighted average of the MHD and PIC outputs as follows:
\begin{align}
    Q_{\text{Int}} = F Q_{\text{MHD}} + (1 - F) Q_{\text{PIC}}, 
    \label{chap2:eq:interlocking}
\end{align}
where $Q_{\text{MHD}}$ and $Q_{\text{PIC}}$ are the physical quantity of MHD and PIC. When calculating $Q_{\text{Int}}$ in MHD domain, $Q_{\text{MHD}}$ is from MHD directly and $Q_{\text{PIC}}$ is calculated from sub-subsection \ref{chap2:conversion_of_physical_quantities_from_pic_to_mhd}. When calculating $Q_{\text{Int}}$ in PIC domain, $Q_{\text{MHD}}$ is calculated from sub-subsection \ref{chap2:conversion_of_physical_quantities_from_mhd_to_pic} and $Q_{\text{PIC}}$ is from PIC directly. Note that the labels of MHD and PIC for $Q$ in this section are different from sub-subsection \ref{chap2:conversion_of_physical_quantities_from_pic_to_mhd} and \ref{chap2:conversion_of_physical_quantities_from_mhd_to_pic}. $F$ is the weight function, and in this work, we employ the following function: 
\begin{align}
    F(y) = \frac{1}{2} \left(1 \pm \cos \left( \frac{y - y_{\text{PIC}}}{y_{\text{MHD}} - y_{\text{PIC}}} \right) \right).
    \label{chap2:eq:interlocking_used_function}
\end{align}
Figure \ref{chap2:fig:multi-hierarchy_scheme_space} shows the schematic picture of multi-hierarchy scheme in space. $y_{\text{MHD}} - y_{\text{PIC}}$ in \eqref{chap2:eq:interlocking_used_function} corresponds to the size of the interface region (green area). Details of how to calculate $Q_{\text{PIC}}$ for MHD and $Q_{\text{MHD}}$ for PIC are explained in the following subsections.

In this study, the interface region ($y_{\text{MHD}} - y_{\text{PIC}}$) is defined to span only a few PIC grid cells. Physical quantities of MHD are sent to PIC as nearly boundary condition (green area), and almost all physical quantities of PIC are sent to MHD (yellow and green areas).

\begin{figure*}[!t]
    \includegraphics[width=\linewidth]{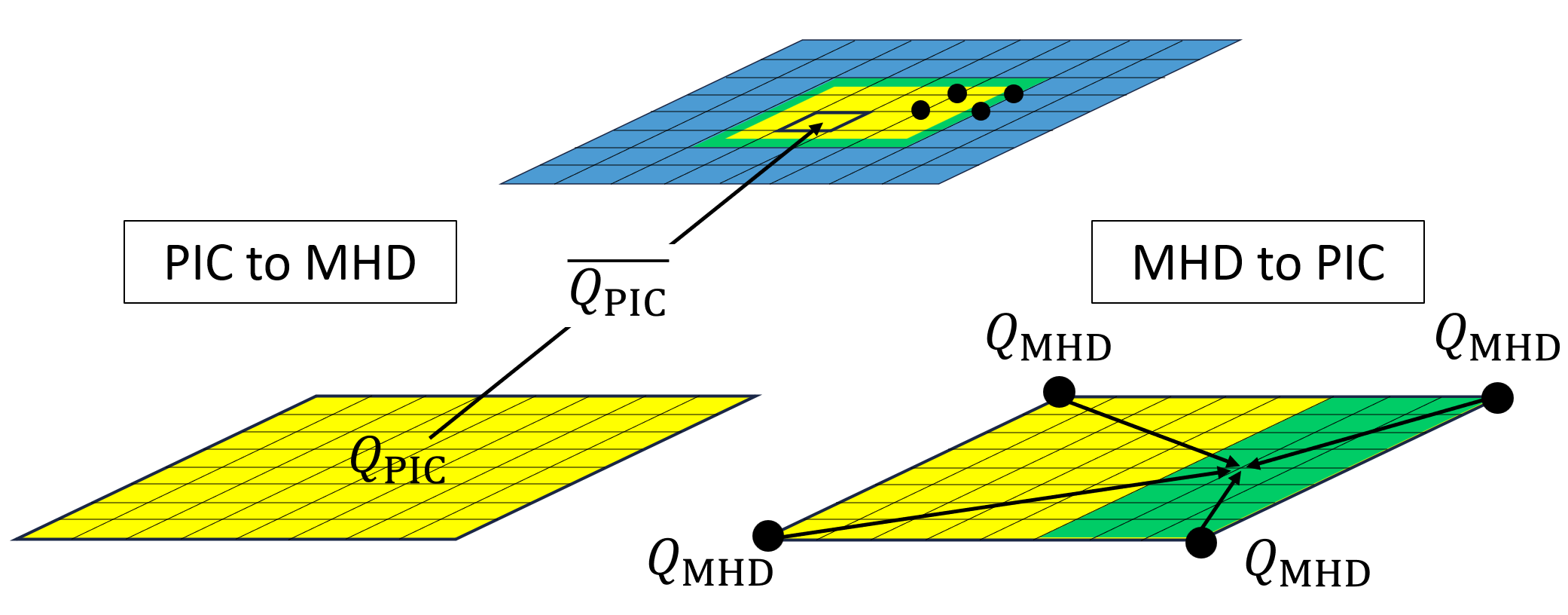}
    \caption{\label{chap2:fig:multi-hierarchy_scheme_information_exchange}Schematic picture of multi-hierarchy simulation scheme of exchanging physical quantities between MHD and PIC. The upper panel shows MHD domain; blue region corresponds to purely MHD region, green region corresponds to interface region where physical quantities of MHD and PIC are mixed, and yellow region corresponds to PIC embedded region. The lower two panels show PIC domain; boundary thick line represents MHD grid and thin line inside represents PIC grid. In this figure $8 \times 8$ PIC grids correspond to 1 MHD grid. Physical quantities of PIC are averaged and sent to MHD, and those of MHD are interpolated and sent to PIC.
    {Alt text: Three figures are shown; one MHD domain and two PIC domains which represent the exchanges of the physical quantities between MHD and PIC.}}
\end{figure*}

\subsubsection{Conversion of physical quantities from PIC to MHD}\label{chap2:conversion_of_physical_quantities_from_pic_to_mhd}

The equations to convert the physical quantities of PIC to MHD are as follows:
\begin{align}
    &n_{i,e,\text{MHD}} = \int f_{i,e} \dd{\bm{v}}, \\
    &\rho_{\text{MHD}} = n_{i,\text{PIC}} m_i + n_{e,\text{PIC}} m_e, \\ 
    &\bm{v}_{i,e,\text{MHD}} = \frac{\int f_{i,e} \bm{v} \dd{\bm{v}}}{n_{i,e,\text{PIC}}}, \\
    &\bm{v}_{\text{MHD}} = \frac{n_{i,\text{PIC}} m_i \bm{v}_{i,\text{PIC}} + n_{e,\text{PIC}} m_e \bm{v}_{e,\text{PIC}}}{\rho_{\mathrm{PIC}}}, \\
    &k_{\rm B} T_{i,e,\text{MHD}} = m_{i,e}  \tr \qty( \int f_{i,e} (\bm{v} - \bm{v}_{i,e,\text{PIC}})(\bm{v} - \bm{v}_{i,e,\text{PIC}}) \dd{\bm{v}} ), \\
    &p_{\text{MHD}} = n_{i,\text{PIC}} T_{i,\text{PIC}} + n_{e,\text{PIC}} T_{e,\text{PIC}}, \\
    &\bm{B}_{\text{MHD}} = \bm{B}_{\text{PIC}}. 
\end{align}

These conversions are applied when the PIC quantities (defined at the PIC grid points) are transferred to MHD domain at each MHD step (shown as yellow and green regions in Figure \ref{chap2:fig:multi-hierarchy_scheme_space}). Physical quantities of PIC are averaged in $\Delta_{\text{MHD}} \times \Delta_{\text{MHD}}$ grids (averaged in 9 grids and sent to 1 MHD grid in Figure \ref{chap2:fig:multi-hierarchy_scheme_space}). Physical quantities of MHD are overwritten by $Q_{\text{Int}}$ calculated using \eqref{chap2:eq:interlocking} in the yellow and green areas. For example, $\rho_{\text{int}} = F \rho_{\text{MHD}} + (1 - F) \overline{\rho_{\text{PIC}}}$ where $\overline{\rho_{\text{PIC}}}$ is averaged mass density in $\Delta_{\text{MHD}} \times \Delta_{\text{MHD}}$ grids. Finally, MHD information where PIC embedded is rewritten as averaged PIC information. Schematic picture for how to exchange physical quantities from PIC to MHD is shown in Figure \ref{chap2:fig:multi-hierarchy_scheme_information_exchange}. 

\subsubsection{Conversion of physical quantities from MHD to PIC}\label{chap2:conversion_of_physical_quantities_from_mhd_to_pic}

The equations to convert the physical quantities of MHD to the PIC are as follows:
\begin{align}
    &n_{i,\text{PIC}} = n_{e,\text{MHD}} = \frac{\rho_{\text{MHD}}}{m_i + m_e}, \\ 
    &\bm{v}_{i,\text{PIC}} = \bm{v}_{\text{MHD}}, \\
    &\bm{v}_{e,\text{PIC}} = \bm{v}_{\text{MHD}} + \frac{\nabla \times \bm{B}_{\text{MHD}}}{\mu_0 q_e n_{e,\text{MHD}}}, \\ 
    &k_{\rm B} T_{i,e,\text{PIC}} = \frac{p_{\text{MHD}}}{2 n_{i,e,\text{MHD}}}, \\
    &\bm{B}_{\text{PIC}} = \bm{B}_{\text{MHD}}, \\
    &\bm{E}_{\text{PIC}} = -\bm{v}_{\text{MHD}} \times \bm{B}_{\text{MHD}}, \\
    &\bm{j}_{\text{PIC}} = \frac{\nabla \times \bm{B}_{\text{MHD}}}{\mu_0}.
\end{align}

These conversions are applied when the MHD quantities (defined at the MHD gird points) are transferred to the PIC domain at each PIC time step in the interface region (shown as green region in Figure \ref{chap2:fig:multi-hierarchy_scheme_space}). Since the MHD grid is much coarser than the PIC grid, we apply the linear interpolation in space to obtain the values on the PIC grids, as illustrated in Figure \ref{chap2:fig:multi-hierarchy_scheme_information_exchange}. Particles inside the interface region are deleted and reloaded as shifted Maxwellian distribution. We assume that the bulk speed is determined by ions, the current is carried only by electrons. This condition is satisfied when the mass ratio $m_i / m_e$ is large enough. The pressures of ions and electrons are assumed to be identical. Schematic picture for how to exchange physical quantities from MHD to PIC is shown in Figure \ref{chap2:fig:multi-hierarchy_scheme_information_exchange}.

Detailed process of particles inside the interface region is as follows:
\begin{enumerate} 
    \item Calculate $n_{\text{Int}}$, $\bm{v}_{\text{Int}}$, and $T_{\text{Int}}$ from Equation \eqref{chap2:eq:interlocking}. Each quantity of PIC is calculated from zeroth, first, and second moments same as equations in sub-subsection \ref{chap2:conversion_of_physical_quantities_from_pic_to_mhd}. How to calculate physical quantities of MHD at each PIC step is explained in sub-subsection \ref{chap2:time_integration}.
    \item Delete particles from computational memory according to the probability of $F$. Random number $rand \in [0, 1)$ is generated for each particle and delete particle if $rand < F(y)$ where $y$ denotes the particle position of y axis.
    \item $n_{\text{Int}}$ number of particles are generated from shifted Maxwellian distribution of $\bm{v}_{\text{Int}}$ and $T_{\text{Int}}$. Position of particle is decided following uniform distribution in each cell.
    \item Load generated particles into computational memory. 
\end{enumerate}
To reduce computational cost, particularly the time for memory allocation and access, we allocate arrays of sufficient size to store particle information and use a boolean variable for each particle to specify which elements are active. This method reduces the overhead of repeatedly releasing and reallocating memory in each PIC step, leading to faster computation.

\subsubsection{Time integration}\label{chap2:time_integration}

\begin{figure}[!t]
    \includegraphics[width=\linewidth]{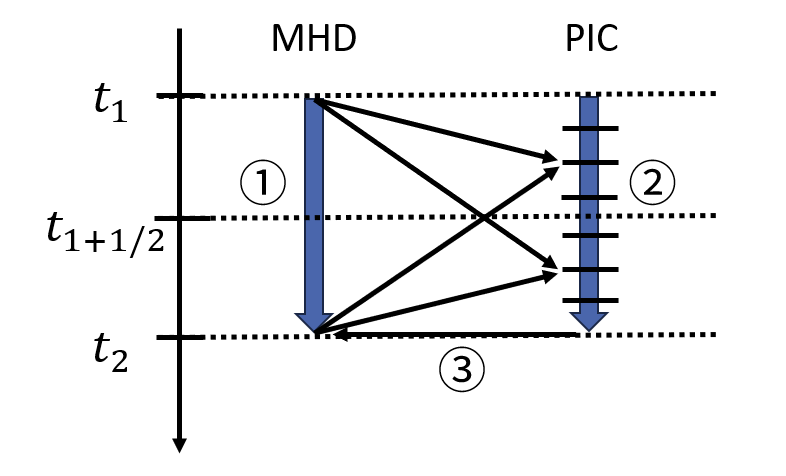}
    \caption{\label{chap2:fig:multi-hierarchy_scheme_time}Schematic picture of multi-hierarchy simulation scheme in time. MHD and PIC are integrated with exchanging physical quantities with each other.
    {Alt text: Schematic picture of multi-hierarchy simulation in time.}}
\end{figure}

Figure \ref{chap2:fig:multi-hierarchy_scheme_time} shows the schematic picture of time integrator in multi-hierarchy scheme. The detail is as follows: 
\begin{enumerate}
    \item Advance MHD by one step with $\Delta t_{\text{MHD}}$ using MHD schemes explained in subsection \ref{chap2:mhd}.
    \item Advance PIC by $\Delta t_{\text{MHD}} / \Delta t_{\text{PIC}}$ steps using PIC schemes explained in subsection \ref{chap2:pic}. In each PIC step, physical quantities of MHD are interpolated and sent to the interface region (green area) in PIC. Sending quantities of MHD are made by linear interpolation of $Q_{\text{MHD}}(t = t_1)$ and $Q_{\text{MHD}}(t = t_2)$.
    \item Send averaged physical quantities of PIC to MHD. 
\end{enumerate}

\subsubsection{Grid size and time step}\label{chap2:grid_size_and_time_step}

The grid size ratio of MHD and PIC $\Delta_{\text{MHD}} / \Delta_{\text{PIC}}$ can be determined freely because MHD model has the spatial and temporal scale-free properties. Since physical quantities are averaged over $\Delta_{\text{MHD}} \times \Delta_{\text{MHD}}$ grids when converting PIC data to MHD data, a larger grid size ratio reduces numerical noise and blocks kinetic effects from propagating into the MHD region. $\Delta_{\text{MHD}}$ should be comparable to the largest typical scales of collisionless plasmas ($\sim$ ion inertial length). Moreover, using coarser MHD grid reduces the computational cost for the projection method. 

The Courant–Friedrichs–Lewy (CFL) condition in PIC calculations is stricter than in MHD calculations because the MHD grid is coarser than the PIC grid. Consequently, we can adopt $\Delta t_{\text{MHD}}$ larger than $\Delta t_{\text{PIC}}$. In low-beta and non-relativistic systems, $\Delta t_{\text{MHD}}$ is estimated as $\Delta_{\text{MHD}}/V_A$, where $V_A$ is the maximum Alfvén speed, while $\Delta t_{\text{PIC}}$ is given by $\Delta_{\text{PIC}}/c$, with $c$ denoting the speed of light. Their ratio is
\begin{align}
    r := \frac{\Delta_{\rm MHD} / V_A}{\Delta_{\rm PIC} / c} \sim \frac{c}{V_{\text{Ai}}} \frac{\Delta_{\rm MHD}}{\Delta_{\rm PIC}} = \frac{\omega_{pe}}{\Omega_{ce}} \sqrt{\frac{m_i}{m_e}} \frac{\Delta_{\rm MHD}}{\Delta_{\rm PIC}}.
\end{align}
$\Delta t_{\text{PIC}}$ and $\Delta t_{\text{MHD}}$ are thus determined as follows: 
\begin{align} 
    &\Delta t_{\text{PIC}} = \min (\text{CFL} \times \Delta_{\text{PIC}} / c, 0.1 \omega_{pe}^{-1}), \\ 
    &\Delta t_{\text{MHD}} = \Delta t_{\text{PIC}} \times [r], 
\end{align} 
where $[r]$ denotes the largest integer not exceeding $r$ (the Gauss symbol). In this paper, we set $\text{CFL} = 0.5$.

\subsubsection{Filtering process}\label{chap2:filter}

In exchanging physical variables between the PIC and MHD domains, numerical noise is often introduced. In the PIC region in particular, the magnetic field defined on the MHD grid frequently exhibits large non-solenoidal noise components. Since we remove the divergence of the magnetic field with the projection method, which solves the Poisson equation, locally large non-solenoidal noise can lead to significant global numerical errors. To mitigate this, we apply a smoothing filter to the variables defined on the MHD grid, specifically given as follows:
\begin{align}
    \overline{Q}_{i, j, \text{MHD}} = \frac{1}{9} \sum_{\alpha, \beta = -1}^{1} Q_{i+\alpha, j+\beta, \text{MHD}}. 
\end{align}
The smoothing filter is used once when the projection method is used. In all simulations in this paper, the projection method and one filtering are employed every $5 - 10$ MHD steps.


\section{Test simulations}\label{chap3:test_simulations}

\subsection{Energy conservation in a static plasma}\label{chap3:energy_conservation}

We first test the conservation of total energy in an unmagnetized plasma, with the initial condition given by
\begin{align*}
    &\rho = n_{{\rm i},0} m_{\rm i} + n_{{\rm e},0} m_{\rm e}, \\
    &u = v = w = 0, \\
    &B_x = B_y = B_z = 0, \\
    &p = p_0 = k_{\rm B} \left( n_{{\rm i},0} T_{{\rm i},0} + n_{{\rm e},0} T_{{\rm e},0} \right).
\end{align*}
Here, the ion and electron masses are denoted by $m_{\rm i}$ and $m_{\rm e}$, respectively, and we set $m_{\rm i}/m_{\rm e} = 25$. The initial number densities of ions ($n_{{\rm i},0}$) and electrons ($n_{{\rm e},0}$) are $n_{{\rm i}, 0} = n_{{\rm e}, 0} = 20$ and $100 \ \text{ppc}$, in order to examine the dependence on the ppc. The ion and electron temperatures are equal ($T_{{\rm i}, 0}/T_{{\rm e}, 0} = 1$), with a thermal velocity of $v_{\rm th} = 0.1c$. The grid size ratio between MHD and PIC is $\Delta_{\text{MHD}} / \Delta_{\text{PIC}} = 10$. The domain size is $N_{x,\text{PIC}} \times N_{y,\text{PIC}} = 100 \times 100$ and $N_{x,\text{MHD}} \times N_{y,\text{MHD}} = 10 \times 100$, corresponding to $100 \times 1000$ in the PIC grid. Symmetric boundaries are imposed in $x$ direction, and periodic boundaries in $y$ direction. For comparison, a PIC simulation with the same setup was also performed, using 100 ppc and periodic boundaries in both directions.

\begin{figure}[!t]
    \includegraphics[width=\linewidth]{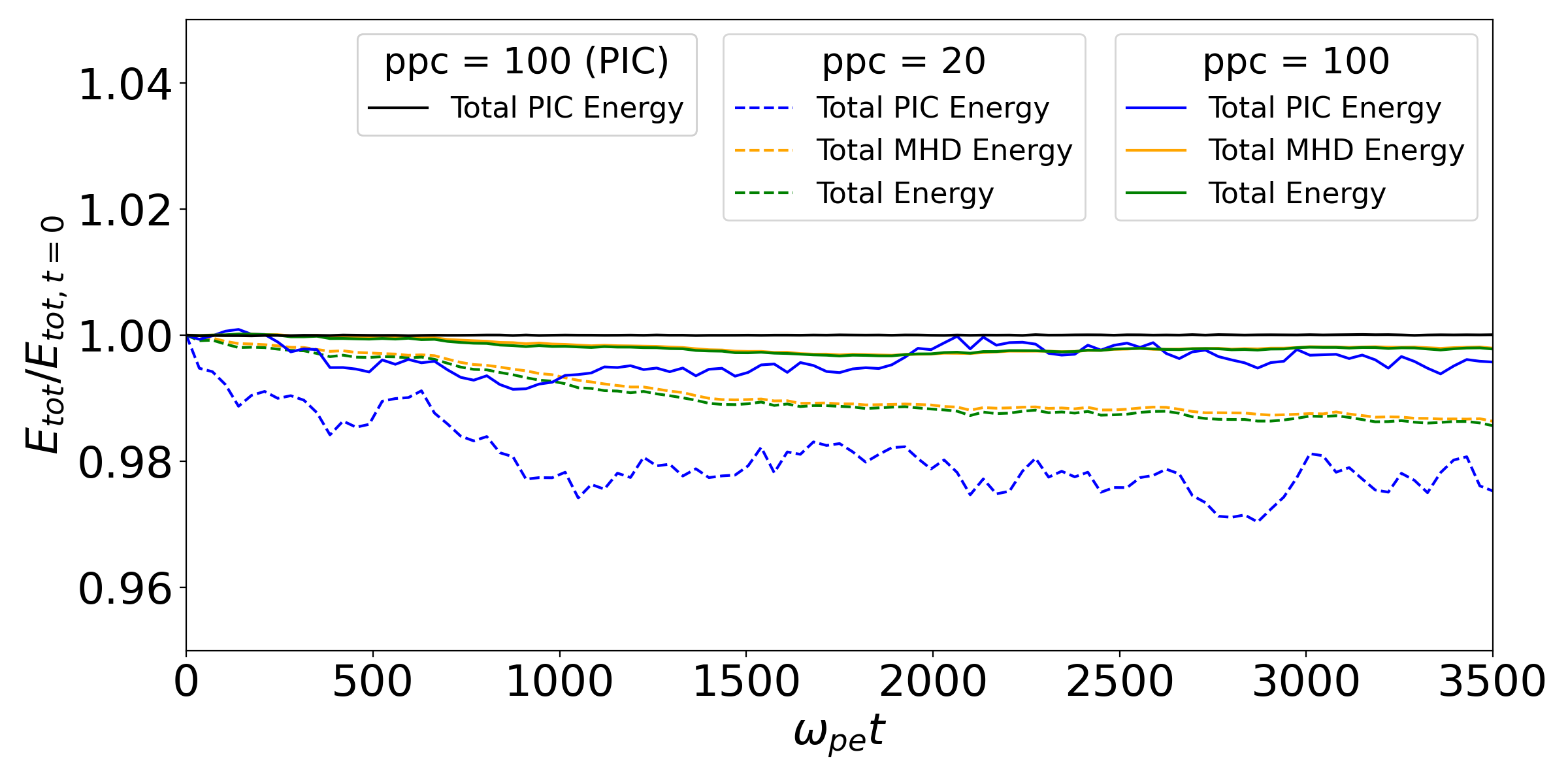}
    \caption{\label{chap3:fig:energy_conservation}Time evolution of the total energy in multi-hierarchy simulations and PIC simulation. The dotted line shows the result of multi-hierarchy simulation with 20 ppc, and the solid line shows the result with 100 ppc. Black line shows the result of PIC simulation with 100 ppc.
    {Alt text: The horizontal axis shows the time $\omega_{pe} t$. The range in horizontal axis is from 0 to 3500. The axis shows the total energy divided by the initial total energy. The range in axis is from 0.95 to 1.05.}}
\end{figure}

Figure \ref{chap3:fig:energy_conservation} presents the time evolution of the total energy in multi-hierarchy simulations with 20 ppc and 100 ppc, together with a PIC simulation at 100 ppc. All simulations solve 50000 PIC steps for time evolution of $3500\omega_{pe} t$. For evolutions exceeding $\mathcal{O}(10^4)$ steps, total energy is conserved within a few $\%$ for 20 ppc and within $0.5 \%$ for 100 ppc. Increasing ppc enhances energy conservation. However, for a fixed ppc, multi-hierarchy simulations show greater energy loss than PIC. As plasma energy dominates the total energy, this loss likely arises from particle deletion and reloading at the interface region. The impact of particle-number fluctuations becomes more pronounced at lower ppc. We recall that particles are first removed with a probability given by $F$ in \eqref{chap2:eq:interlocking}, and then reloaded as $n_{\text{Int}}$, rounded to the nearest integer in the interface region.

\begin{figure}[!t]
    \includegraphics[width=\linewidth]{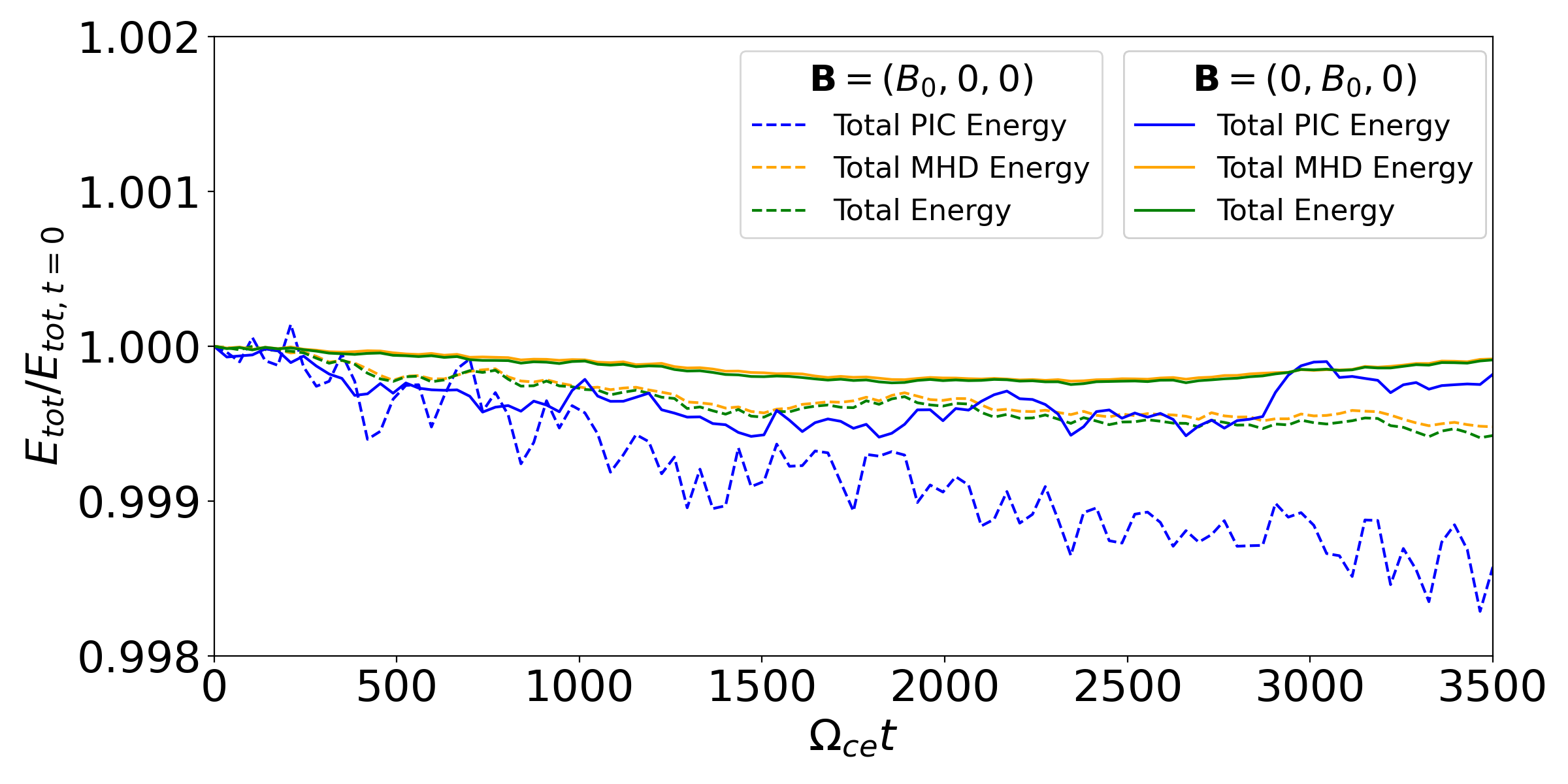}
    \caption{\label{chap3:fig:energy_conservation_bx_and_by}Time evolution of the total energy in multi-hierarchy simulations with magnetic field. 100 ppc is used in both simulations.
    {Alt text: The axis shows the time $\Omega_{ce} t$. The range in axis is from 0 to 3500. The axis shows the total energy divided by the initial total energy. The range in axis is from 0.998 to 1.002.}}
\end{figure}

We verify the conservation of total energy in a static magnetized plasma. A uniform magnetic field is added in the $x$ or $y$ direction, i.e. $B_x = B_0$ or $B_y = B_0$, where $B_0$ is set by $\omega_{pe} / \Omega_{ce} = 1$ (corresponding to the plasma beta of 0.04). All other parameters are identical to the previous simulation, with 100 ppc used. Figure \ref{chap3:fig:energy_conservation_bx_and_by} presents the time evolution of the total energy. The total energy is conserved within $0.2 \%$, which is better than the case without a magnetic field. In these runs, the magnetic energy dominates over the plasma energy, making the errors from particle deletion and reloading insignificant. The total energy is particularly well conserved when $B_y = B_0$, suggesting that conservation is improved when the particle gyration plane is parallel, rather than perpendicular, to the interface region.

\begin{figure}[!t]
    \includegraphics[width=\linewidth]{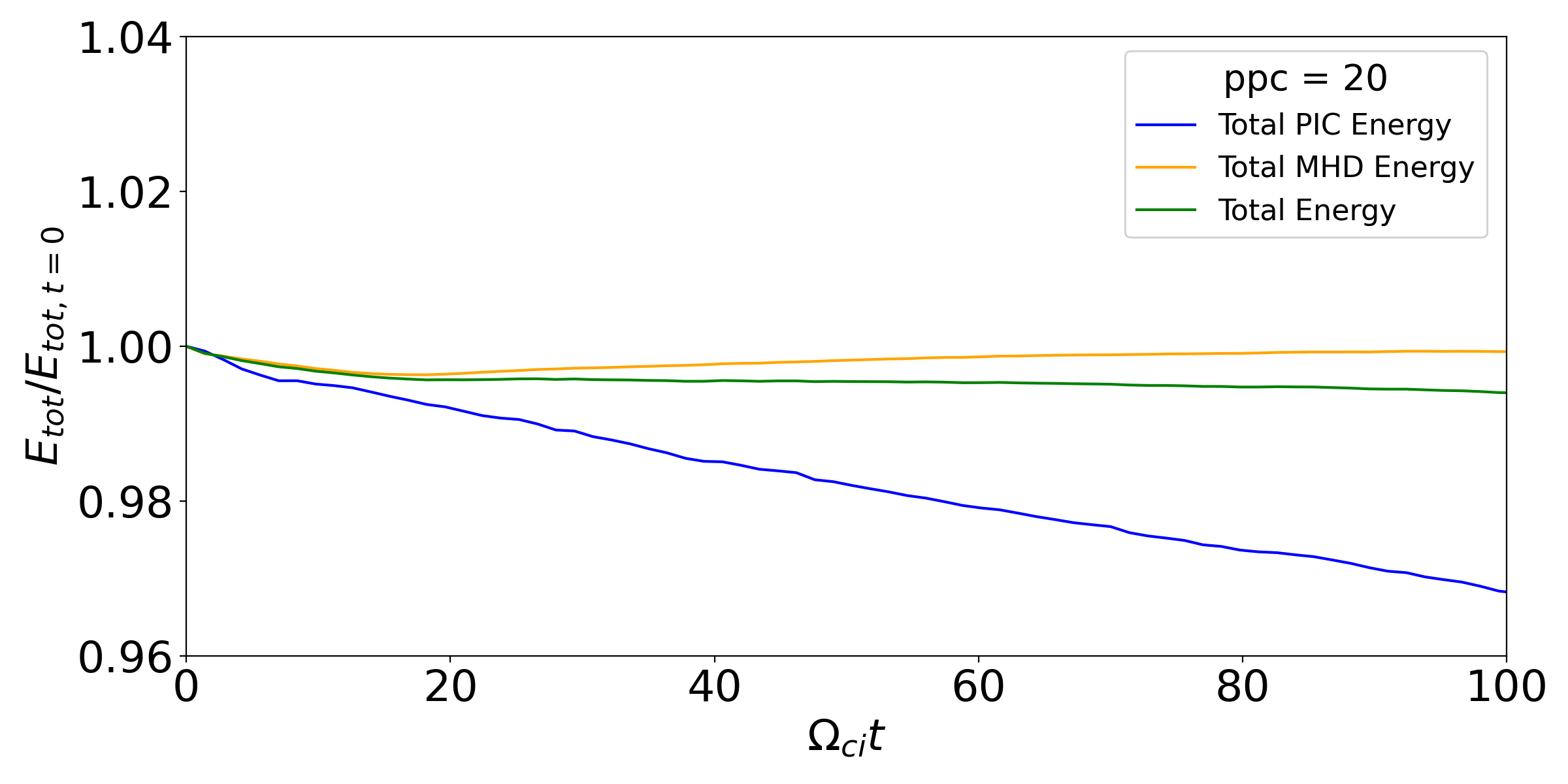}
    \caption{\label{chap3:fig:energy_conservation_forcefree}Time evolution of the total energy in multi-hierarchy simulations with forcefree current sheet. 20 ppc is used in this simulation.
    {Alt text: The axis shows the time $\Omega_{ci} t$. The range in axis is from 0 to 100. The axis shows the total energy divided by the initial total energy. The range in axis is from 0.96 to 1.04.}}
\end{figure}

We also verify the time evolution of the total energy under the initial conditions of a force-free current sheet following \citet{makwana2017}, who demonstrated that the total energy is conserved within $0.1 \%$ when using 1000 ppc in their multi-hierarchy model. The simulation setup is as follows:
\begin{align*}
    &\rho = \rho_0, \\
    &u= 0, \\
    &v= 0, \\ 
    &w= 0, \\
    &B_x = B_0 \tanh(y / \delta),\\
    &B_y = 0,\\
    &B_z = B_0 / \cosh(y / \delta),\\ 
    &p = p_0. 
\end{align*}
We set $m_{\rm i} / m_{\rm e} = 25$, $n_{{\rm i}, 0} = n_{{\rm e}, 0} = 20 \ \text{ppc}$, $T_{{\rm i}, 0}/T_{{\rm e}, 0} = 1$, and $\omega_{pe} / \Omega_{ce} = 1$. The plasma beta is set to 0.25. Grid-size ratio is adopted as $\Delta_{\text{MHD}} / \Delta_{\text{PIC}} = 10$. The size of the simulation domain is $2000 \times 200$ for the PIC region and $200 \times 100$ for the MHD region (corresponding to $2000 \times 1000$ PIC grids). It corresponds to $100\lambda_i \times 50\lambda_i$ simulation box. The current sheet thickness $\delta$ is set to $1\lambda_i$. Periodic boundary condition is imposed in the $x$ direction, while symmetric boundary condition is imposed in the $y$ direction.

Figure \ref{chap3:fig:energy_conservation_forcefree} presents the time evolution of the total energy with forcefree current sheet. Our results are within $0.5\%$ when 20 ppc is used after $100 \Omega_{ci} t$ has passed. Previous simulation results in Figure \ref{chap3:fig:energy_conservation} indicate that using 5 times ppc reduces the energy conservation error by at least half. The energy conservation property is consistent with the result of \citet{makwana2017}.

\subsection{Alfv\'en wave propagation}\label{chap3:alfven_wave_propagation}

To verify that physical quantities are correctly exchanged between the MHD and PIC regions, we next perform a test of circularly-polarized Alfv\'en wave propagation. The initial conditions are as follows.
\begin{align*}
    &\rho = n_{{\rm i},0} m_{\rm i} + n_{{\rm e},0} m_{\rm e}, \\
    &u = \delta V_A \sin(k y), \\
    &v = 0, \\ 
    &w = \delta V_A \cos(k y), \\
    &B_x = -\delta B_0 \sin(k y), \\
    &B_y = B_0, \\ 
    &B_z = -\delta B_0 \cos(k y), \\ 
    &p = p_0 = k_{\rm B} \left( n_{{\rm i},0} T_{{\rm i},0} + n_{{\rm e},0} T_{{\rm e},0} \right).
\end{align*}
We set $m_{\rm i}/m_{\rm e} = 25$, $n_{{\rm i},0} = n_{{\rm e},0} = 50 \ \text{ppc}$, $T_{{\rm i}, 0}/T_{{\rm e}, 0} = 1$ (with the thermal speed of $v_{\rm the} = 0.1c$), and $\omega_{pe}/\Omega_{ce} = 1$. $V_A$ denotes the Alfv\'en speed defined by $V_A := B_0 / \sqrt{\mu_0 \rho_0}$. The grid-size ratio is adopted as $\Delta_{\text{MHD}} / \Delta_{\text{PIC}} = 5$. The size of the simulation domain is $25 \times 100$ for the PIC region and $5 \times 1000$ for the MHD region (corresponding to $25 \times 5000$ PIC grids). The ion inertial length $\lambda_i$ is resolved by 50 PIC grids. We use $\lambda = 20\lambda_i$ $(=2\pi/k)$ and $\delta = 0.05$. Symmetric boundaries are imposed in $x$ direction, and periodic boundaries in $y$ direction.

\begin{figure}[!t]
    \includegraphics[width=\linewidth]{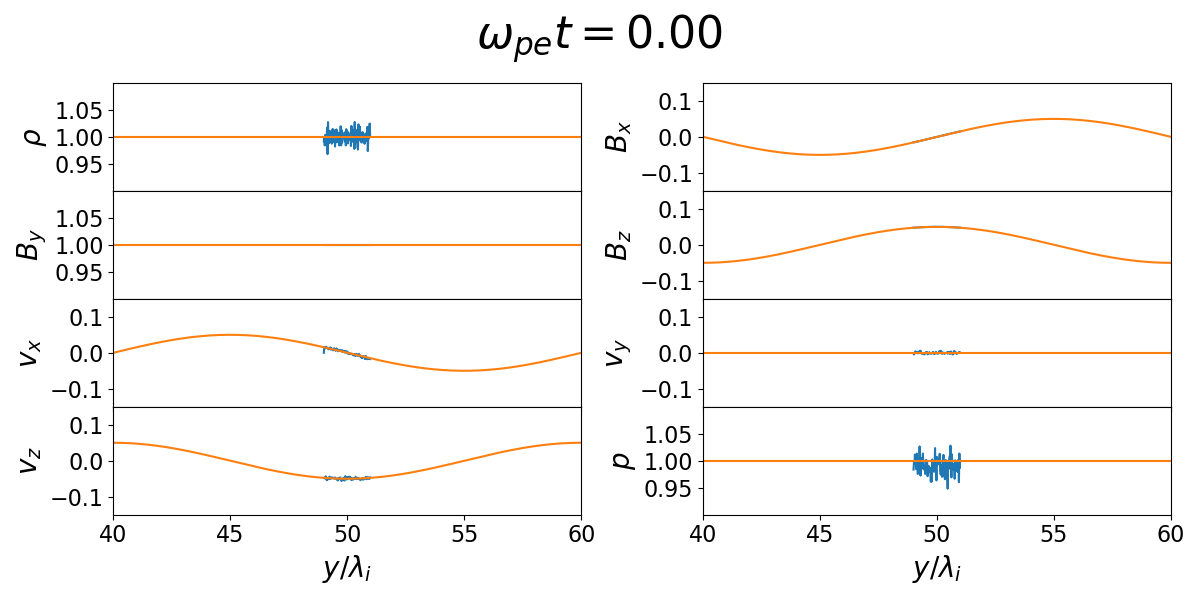}
    \includegraphics[width=\linewidth]{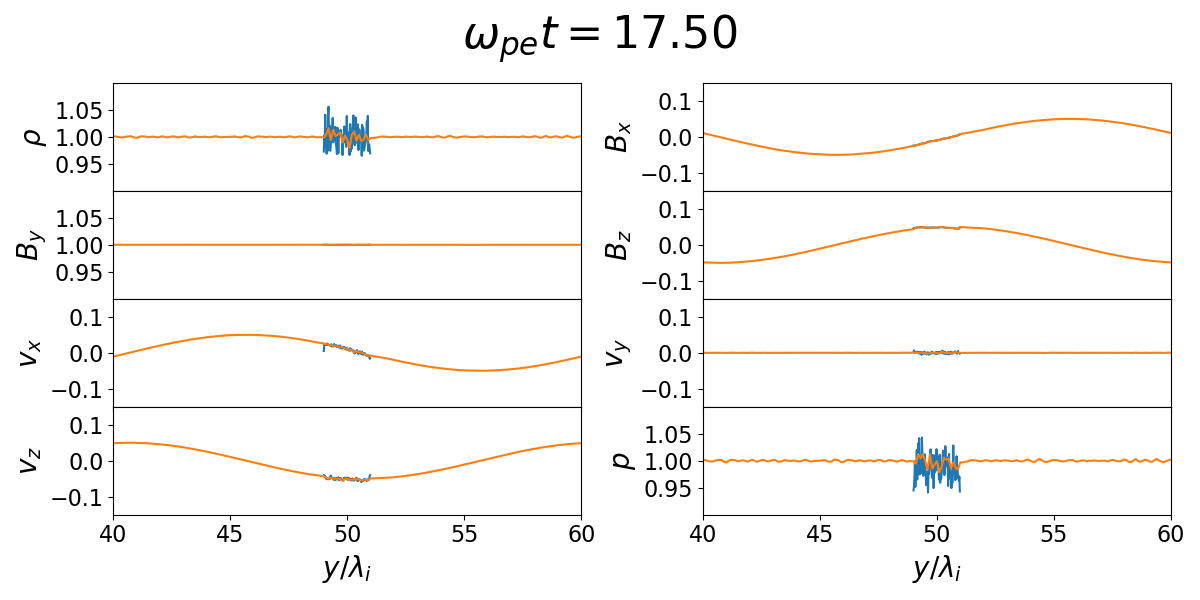}
    \includegraphics[width=\linewidth]{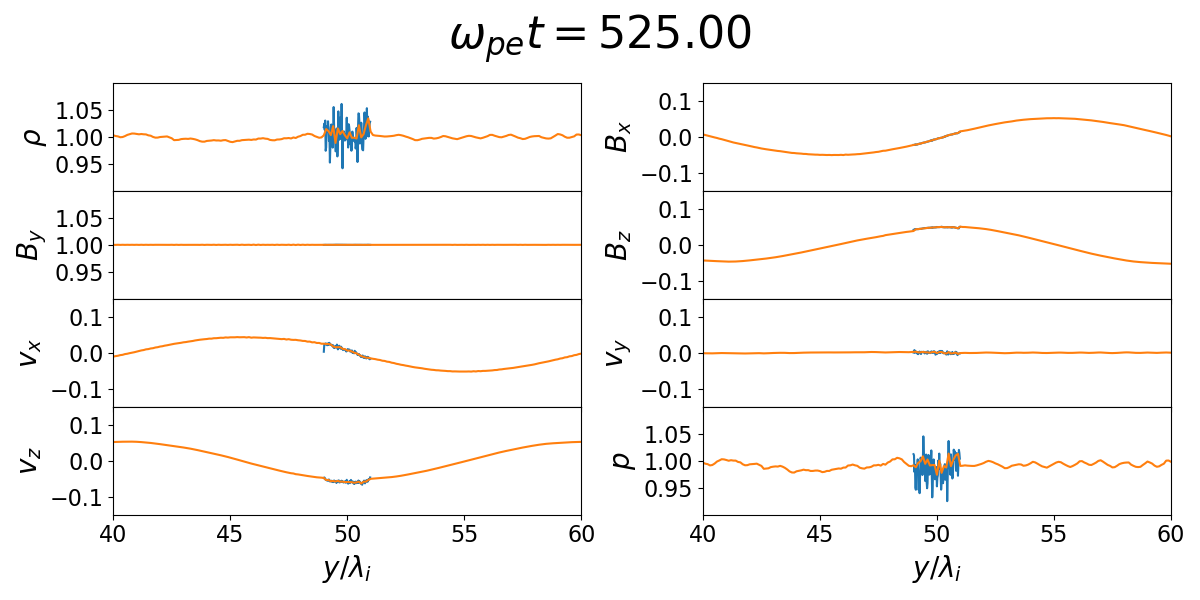}
    \caption{\label{chap3:fig:alfven}Snapshots of Alfv\'en wave propagation at $\omega_{pe} t = 0.00, 17.50, 525.00$. $\rho, B_x, B_y, B_z, V_x, V_y, V_z$, and $p$ are plotted from the top left. Blue line shows the result obtained from PIC, and orange line shows the result obtained from MHD.
    {Alt text: The axis shows the position normalized by $\lambda_i$. The range in axis is from 40 to 60. The axis shows the amplitude of each physical quantity. The range in axis is about $\pm 0.1$ from the initial background value.}}
\end{figure}

Figure \ref{chap3:fig:alfven} presents snapshots of Alfv\'en wave propagation at $\omega_{pe} t = 0.00, 17.50, 525.00$. At $\omega_{pe} t = 17.50$, small fluctuations appear in $\rho$ and $p$ throughout the simulation domain. These fluctuations are not introduced from the PIC region; rather, they are generated simultaneously across the entire MHD domain. They are most likely caused by numerical errors in the projection method (described in sub-subsection \ref{chap2:filter}), which affects $\rho$ and $p$. Because the vertical axis range of $\rho$ and $p$ is small, these weak fluctuations appear more prominently. At $\omega_{pe} t = 525.00$, after one wavelength has passed through the PIC region, the Alfv\'en wave retains its shape in both the magnetic field and flow velocity, confirming its successful propagation through the PIC region.

\begin{figure}[!t]
    \includegraphics[width=\linewidth]{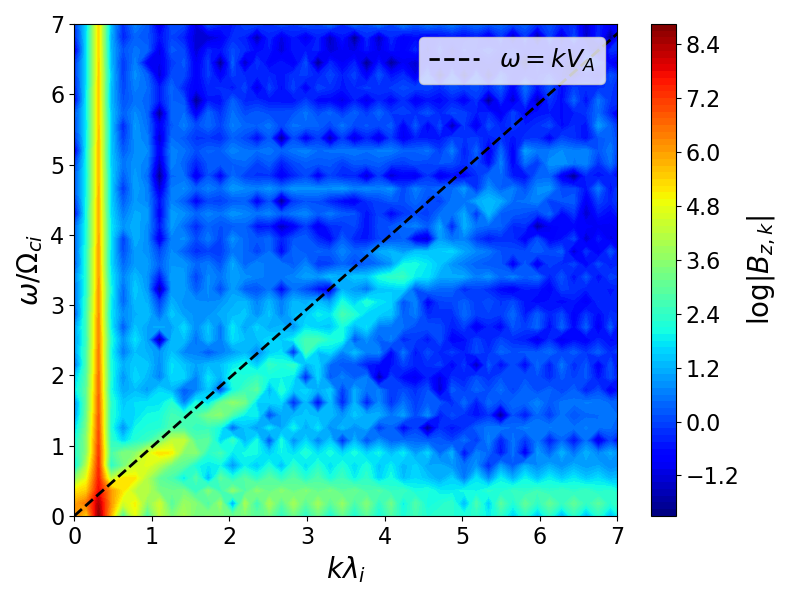}
    \caption{\label{chap3:fig:alfven_dispersion}$\omega - k$ diagram constructed using the MHD's $B_z$ data from the multi-hierarchy simulation of Alfv\'en wave propagation. The wavenumber is normalized by the ion inertial length $\lambda_i$, and the frequency is normalized by the ion gyro frequency $\Omega_{ci}$. 
    {Alt text: The axis shows the $k \lambda_i$. The range in axis is from 0 to 7. The axis shows the $\omega / \Omega_{ci}$. The range in axis is from 0 to 7. The contour of $\log |B_{z, k}|$ is plotted in the range from -2.0 to 9.0.}}
\end{figure}

Figure \ref{chap3:fig:alfven_dispersion} shows the $\omega - k$ diagram constructed from the $B_z$ of MHD. We used $B_z$ within $y = [30\lambda_i, 70\lambda_i]$ which corresponds to $2\lambda$ where $\lambda$ is the wavelength of the initial Alfv\'en wave. The prominent peak around $k\lambda_i \sim 0.25$ corresponds to the Alfv\'en wave introduced as the initial condition. In addition, mode along $\omega = k V_A$ is observed. Within the range $\omega < \Omega_{ci}$, Alfv\'en waves entering PIC region are converted into dispersive modes of whistler and ion cyclotron modes, which then converted into the dispersive Alfv\'en mode in MHD region. Modes which are not included in ideal MHD, such as whistler wave, typically treated as numerical noises in ideal MHD, and they can create artificial fluctuation when transmitted into MHD region. The simulation result indicates that the multi-hierarchy scheme proposed in subsection \ref{chap2:multi-hierarchy_scheme} exhibits robustness against numerical noise of PIC when PIC information is sent to MHD. A small peak is also seen for $\omega > \Omega_{ci}$, which should be attributed solely to the conversion of Whistler modes. Our result is consistent with Figures 4 and 5 in \citet{sugiyama2007}; dispersive whistler wave is seen clearly in the PIC region and dispersive Alfv\'en wave is seen slightly in the MHD region.

\begin{figure}[!t]
    \makebox[\linewidth][c]{
        \includegraphics[width=\linewidth]{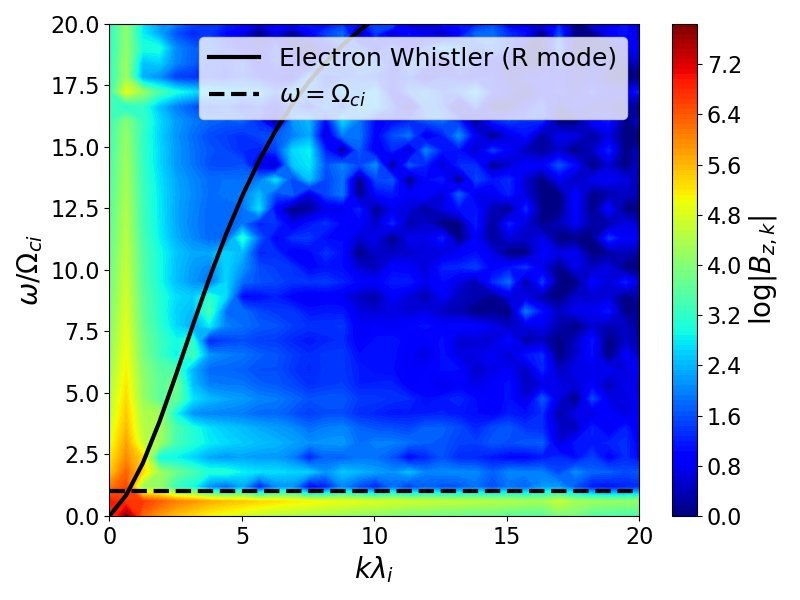}
    }
    \makebox[\linewidth][c]{
        \hspace{6mm}
        \includegraphics[width=\linewidth]{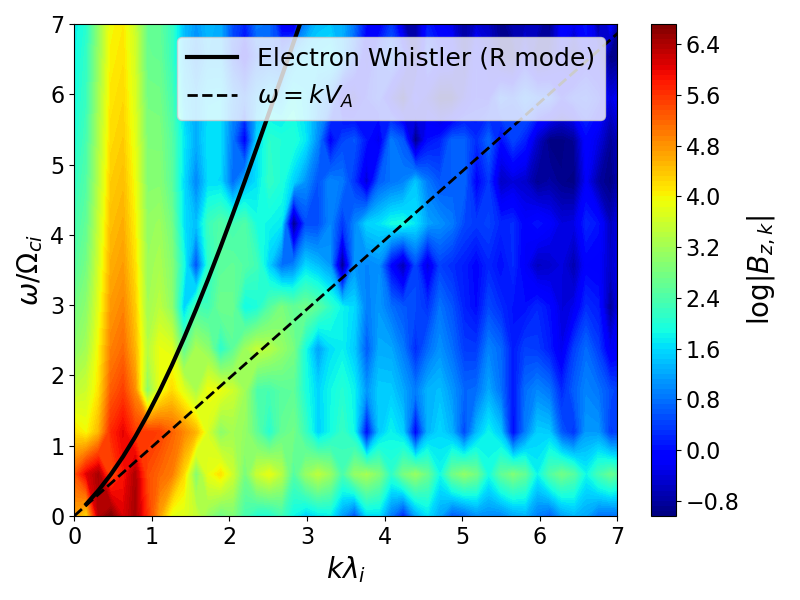}
    }
    \caption{\label{chap3:fig:alfven_dispersion_pic_and_mhd_small}$\omega - k$ diagram from the test simulation of Alfv\'en wave propagation with larger PIC region (1 wavelength). Upper diagram is constructed using $B_z$ in PIC, and the lower diagram is constructed using $B_z$ in MHD. The wavenumber is normalized by the ion inertial length $\lambda_i$, and the frequency is normalized by the ion gyro frequency $\Omega_{ci}$. 
    {Alt text: (Upper panel) The axis shows the $k \lambda_i$. The range in axis is from 0 to 20. The axis shows the $\omega / \Omega_{ci}$. The range in axis is from 0 to 20. The contour of $\log |B_{z, k}|$ is plotted in the range from 0.0 to 8.0. (Lower panel) The axis shows the $k \lambda_i$. The range in axis is from 0 to 7. The axis shows the $\omega / \Omega_{ci}$. The range in axis is from 0 to 7. The contour of $\log |B_{z, k}|$ is plotted in the range from -1.0 to 7.0.}}
\end{figure}

Although unphysical wave-mode conversion is negligible when the wavelength greatly exceeds the PIC domain, it can still occur when the two scales are comparable. To test this, we repeated the calculation with an enlarged PIC region and a shorter wavelength ($\lambda = 10\lambda_i$). The PIC region was set to one wavelength ($N_{y, \text{PIC}} = 500$). The upper panel of Figure \ref{chap3:fig:alfven_dispersion_pic_and_mhd_small} shows the $\omega - k$ spectrum from $B_z$ in the PIC grid, where both a distinct whistler branch (shown by solid line) and the ion-cyclotron branch (near $\omega = \Omega_{ci}$ shown by dotted line) are visible, confirming the conversion to kinetic dispersive waves. The lower panel, constructed from $B_z$ in the MHD grid, shows the existence of Alfv\'en mode along with a whistler-like mode. These results indicate that for short-wavelength waves influenced by kinetic effects, or for waves spanning roughly one wavelength in the PIC region, modes absent in ideal MHD can enter the MHD domain.

\begin{figure}[!t]
    \includegraphics[width=\linewidth]{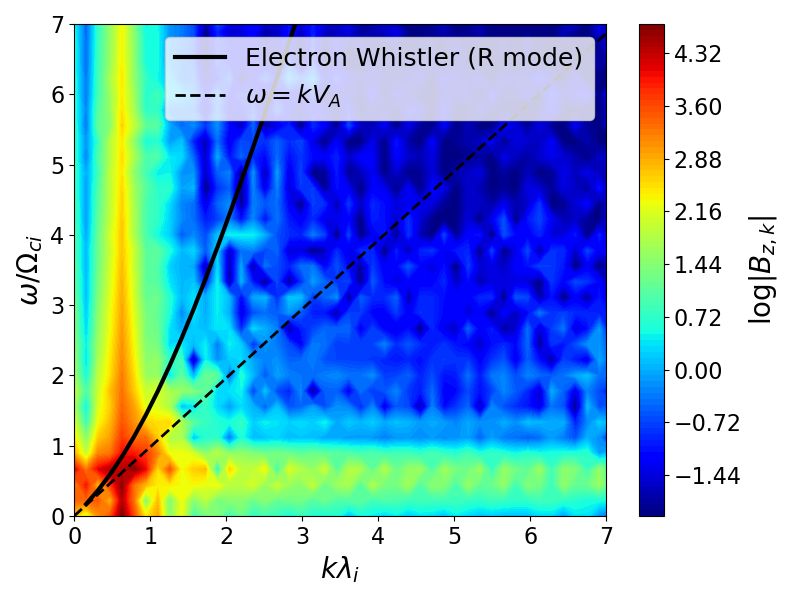}
    \caption{\label{chap3:fig:alfven_dispersion_mhd_coarse} Same as Figure \ref{chap3:fig:alfven_dispersion_pic_and_mhd_small} but with a larger grid-size ratio between the MHD and PIC domains; $\Delta_{\text{MHD}} / \Delta_{\text{PIC}} = 20$ ($\Delta_{\text{MHD}} = 0.4\lambda_i$).
    {Alt text: The axis shows the $k \lambda_i$. The range in axis is from 0 to 7. The axis shows the $\omega / \Omega_{ci}$. The range in axis is from 0 to 7. The contour of $\log |B_{z, k}|$ is plotted in the range from -2.0 to 5.0.}}
\end{figure}

Whistler wave does not exist in ideal MHD so that it should be treated as numerical noises and create unphysical results. This problem may be solved by sufficiently increasing the grid size of MHD. To investigate the dependence on the grid size ratio between MHD and PIC $\Delta_{\text{MHD}} / \Delta_{\text{PIC}}$, $\Delta_{\text{MHD}} / \Delta_{\text{PIC}} = 20$ is adopted (4 times larger). Other parameters are kept the same as in the previous simulation. Here, $\Delta_{\text{MHD}}$ corresponds to $0.4\lambda_i$, where $\lambda_i$ is the ion inertial length. Figure \ref{chap3:fig:alfven_dispersion_mhd_coarse} shows the $\omega - k$ diagram constructed from the $B_z$ in MHD. Compared to the upper panel in Figure \ref{chap3:fig:alfven_dispersion_pic_and_mhd_small}, the peak corresponding to the whistler mode disappears. This result indicates that using sufficiently large MHD grid size makes possible to prevent modes which are absent in ideal MHD from propagating into MHD.

\subsection{Fast mode propagation}\label{chap3:fast_mode_propagation}

Third test simulation setup is the fast mode wave. Solving the propagation of compressive wave needs to handle variation of particles correctly. This test simulation can investigate the robustness for larger numerical noises from particles compared to transverse wave such as previous Alfv\'en wave simulations. Moreover it is useful for checking whether inflow and outflow can be handled when applied to magnetic reconnection. Initial condition is same as previous studies \citep{daldorff2014, makwana2017, haahr2025} and is as follows:
\begin{align*}
    &\rho = \rho_0 (1 + \delta \sin(k y)), \\
    &u = 0, \\
    &v = \delta C_f \sin(k y), \\ 
    &w = 0, \\
    &B_x = 0, \\
    &B_y = 0, \\ 
    &B_z = B_0 (1.0 + \delta \sin(k y)), \\ 
    &p = p_0 (1.0 + \gamma \delta \sin(k y)), 
\end{align*}
We set $m_{\rm i} / m_{\rm e} = 25$, $n_{{\rm i}, 0} = n_{{\rm e}, 0} = 50 \ \text{ppc}$, $T_{{\rm i}, 0}/T_{{\rm e}, 0} = 1$ (with the thermal speed of $v_{\rm the} = 0.1c$), and $\omega_{pe} / \Omega_{ce} = 1$. $C_A$ denotes the fast mode wave speed by $C_A := \sqrt{V_A^2 + C_s^2}$, here $V_A, C_s$ denote the Alfv\'en speed and the sound speed defined by $V_A := B_0 / \sqrt{\mu_0 \rho_0}, C_s := \sqrt{\gamma p_0 / \rho_0}$. The grid-size ratio is adopted as $\Delta_{\text{MHD}} / \Delta_{\text{PIC}} = 5$. The size of the simulation domain is $25 \times 100$ for the PIC region and $5 \times 1000$ for the MHD region (corresponding to $25 \times 5000$ PIC grids). The ion inertial length $\lambda_i$ is resolved 50 PIC grids. We use $\lambda = 2 \pi / k = 20\lambda_i$ and $\delta = 0.05$. Symmetric boundaries are imposed in $x$ direction, and periodic boundaries in $y$ direction.

\begin{figure}[!t]
    \includegraphics[width=\linewidth]{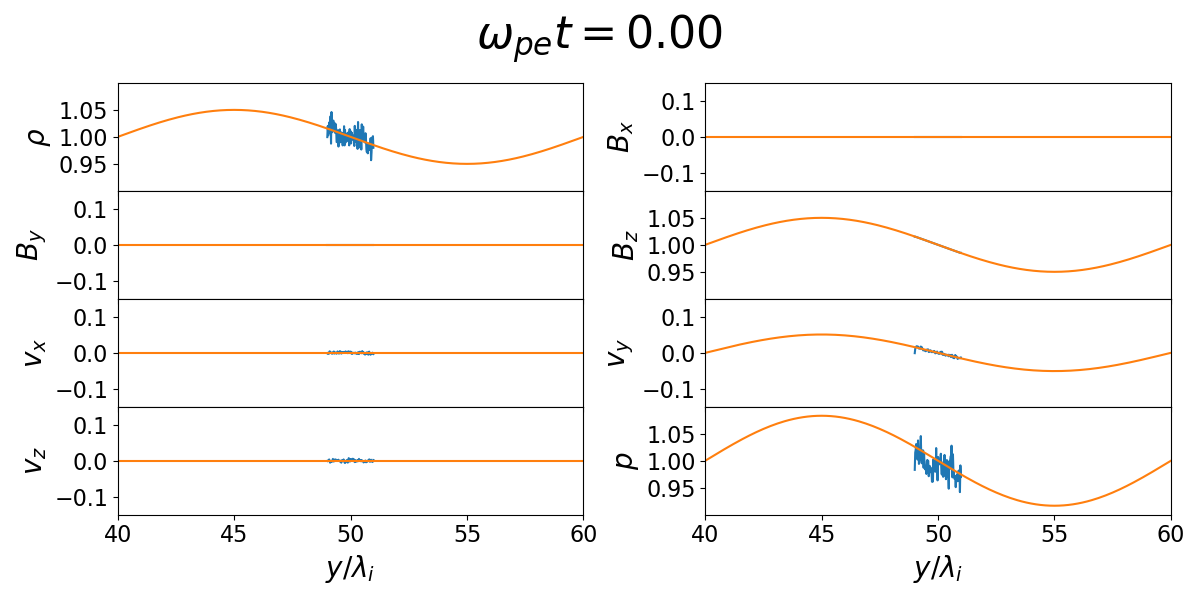}
    \includegraphics[width=\linewidth]{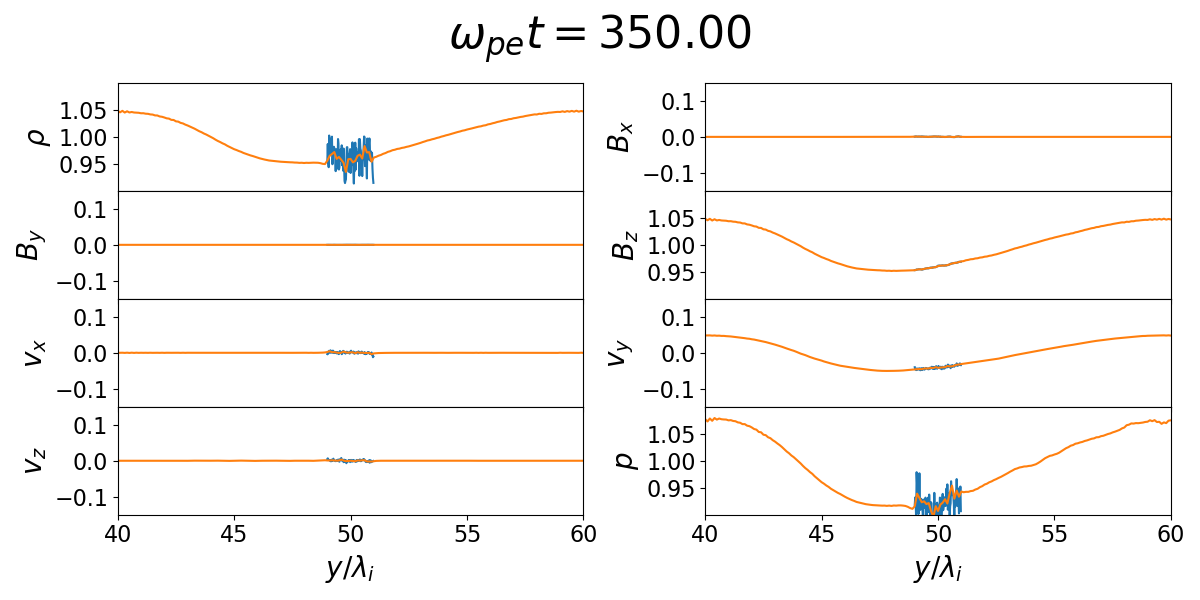}
    \includegraphics[width=\linewidth]{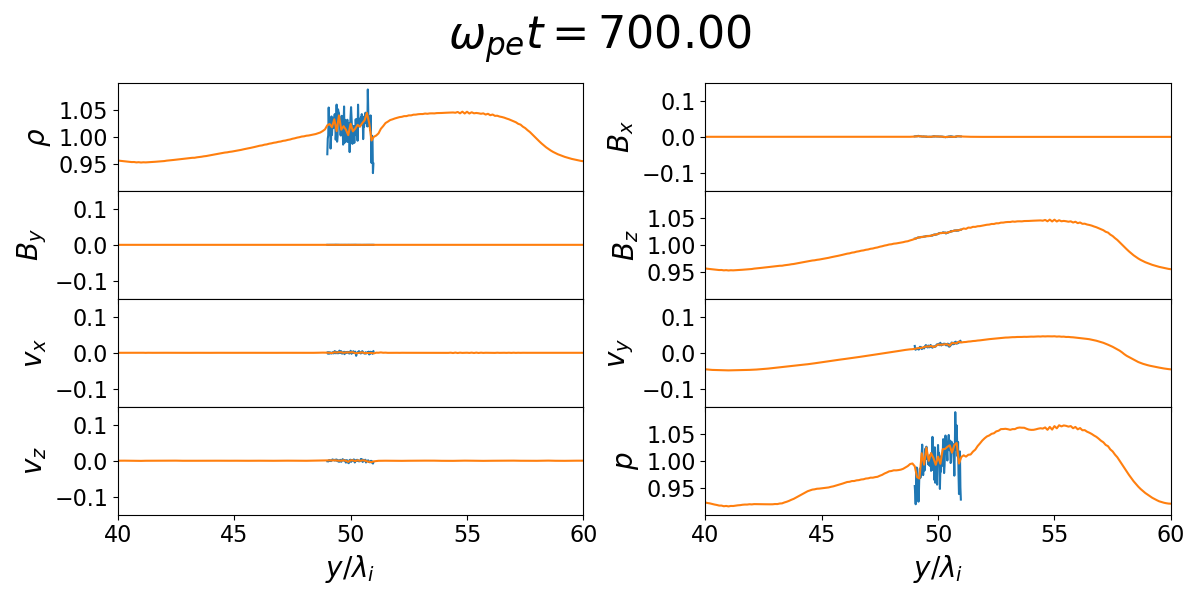}
    \caption{\label{chap3:fig:fast_mode}Snapshots of fast mode wave propagation at $\omega_{pe} t = 0.00, 350.00, 700.00$. $\rho, B_x, B_y, B_z, V_x, V_y, V_z$, and $p$ are plotted from the top left. Blue line shows the result obtained from PIC, and orange line shows the result obtained from MHD.
    {Alt text: The axis shows the position normalized by $\lambda_i$. The range in axis is from 40 to 60. The axis shows the amplitude of each physical quantity. The range in axis is about $\pm 0.1$ from the initial background value.}}
\end{figure}

Figure \ref{chap3:fig:fast_mode} shows the snapshots of fast mode wave propagation at $\omega_{pe} t = 0.00, 350.00, 700.00$. The fast-mode wave steepens during propagation due to nonlinear effects, and its signature appears at $\omega_{pe} t = 350.00$ and $700.00$. The steepened structure located at $y = [40\lambda_i, 45\lambda_i]$ at $\omega_{pe} t = 350.00$ passes through the PIC region, and moves to $y = [55\lambda_i, 60\lambda_i]$ at $\omega_{pe} t = 700.00$, while maintaining waveform. However, the results around PIC region at $\omega_{pe} t = 700.00$, the waveforms of $\rho, p$ are distorted. However, from the results of $B_z$ and $V_y$, multi-hierarchy scheme can propagate compressive wave without distorting its waveform.

\begin{figure}[!t]
    \includegraphics[width=\linewidth]{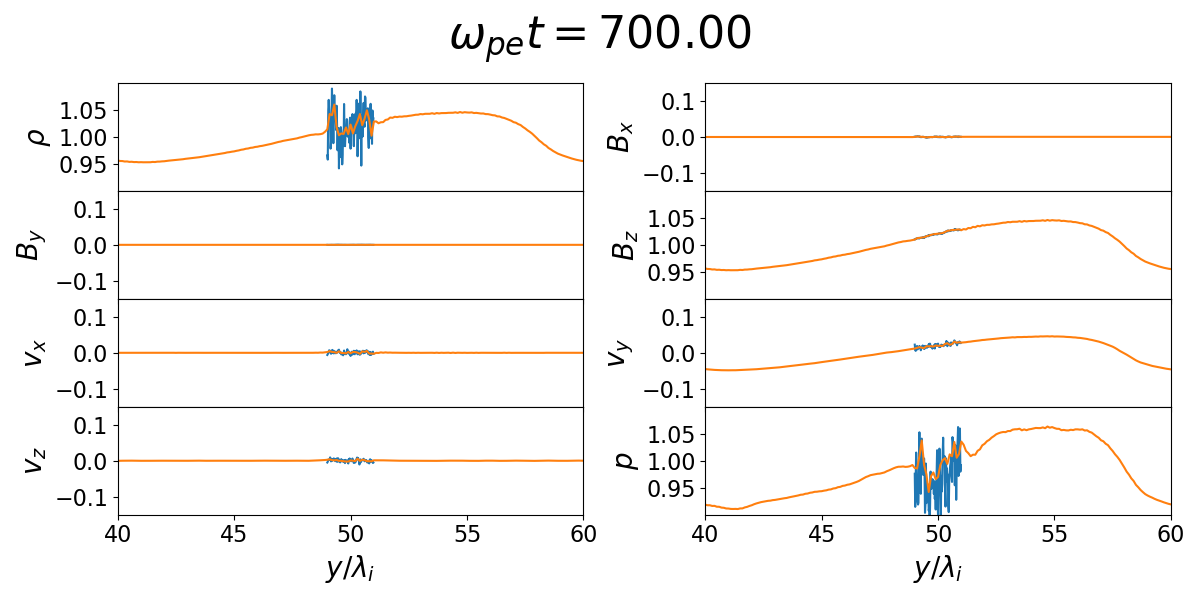}
    \caption{\label{chap3:fig:fast_mode_less_particle}Snapshots of fast mode wave propagation at $\omega_{pe} t = 700.00$. $\rho, B_x, B_y, B_z, V_x, V_y, V_z$, and $p$ are plotted from the top left. Blue line shows the result obtained from PIC, and orange line shows the result obtained from MHD.
    {Alt text: The axis shows the position normalized by $\lambda_i$. The range in axis is from 40 to 60. The axis shows the amplitude of each physical quantity. The range in axis is about $\pm 0.1$ from the initial background value.}}
\end{figure}

To clarify the reason why $\rho$ and $p$ are distorted, we first repeated calculation which changes the number of particles to investigate the effects from numerical noises. Figure \ref{chap3:fig:fast_mode_less_particle} presents the results at $\omega_{pe} t = 700.00$ when using 20 ppc instead of 50 ppc. $\rho$ and $p$ are distorted around the PIC region, while the waveform of $V_y$ and $B_z$ remain undistorted. The results remain largely unchanged compared to previous results even with fewer particles, demonstrating that our multi-hierarchy method is robust to numerical noises from the number of particles. 

\begin{figure}[!t]
    \includegraphics[width=\linewidth]{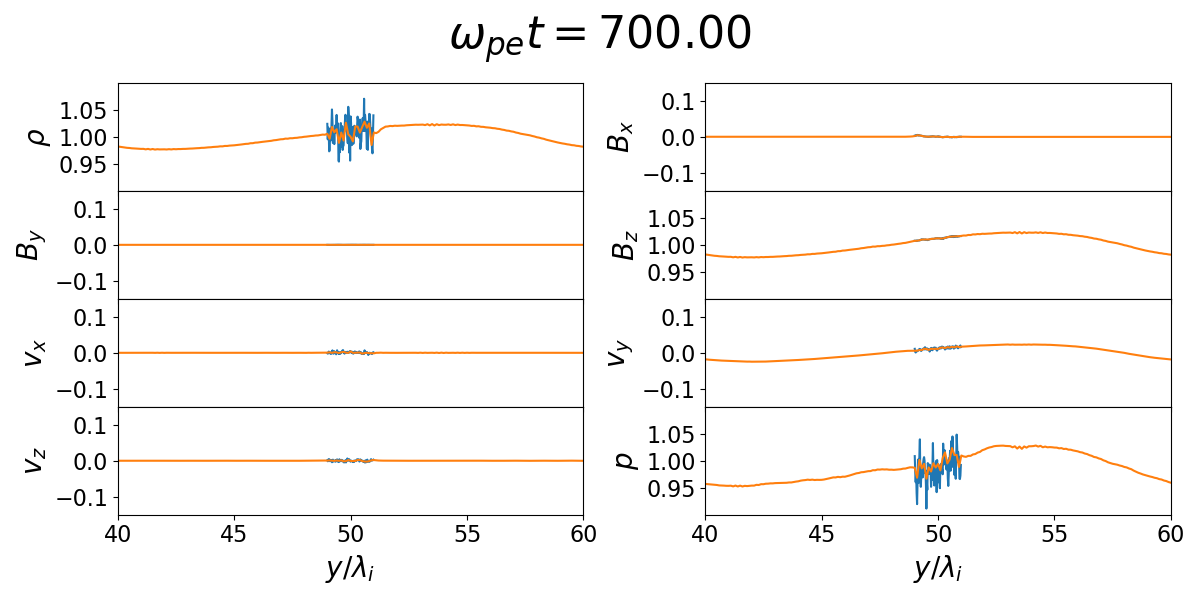}
    \caption{\label{chap3:fig:fast_mode_weak}Snapshots of fast mode wave propagation at $\omega_{pe} t = 700.00$. $\rho, B_x, B_y, B_z, V_x, V_y, V_z$, and $p$ are plotted from the top left. Blue line shows the result obtained from PIC, and orange line shows the result obtained from MHD.
    {Alt text: The axis shows the position normalized by $\lambda_i$. The range in axis is from 40 to 60. The axis shows the amplitude of each physical quantity. The range in axis is about $\pm 0.1$ from the initial background value.}}
\end{figure}

We next repeated calculation which changes the amplitude of the initial wave. Figure \ref{chap3:fig:fast_mode_weak} presents the results at $\omega_{pe} t = 700.00$ when using $\delta = 0.025$ instead of $0.05$. Compared to previous results, $\rho$ and $p$ are not largely distorted. It indicates that the reason of distortion stems from the steepening, which is a nonlinear effect of large amplitude compressive waves. 

\subsection{Shock tube problem}\label{chap3:shock_tube_problem}

Fourth test simulation setup is a shock tube problem. This test simulation verifies the behavior around discontinuity or shock, for example they can be seen in the nonlinear regime of compressive wave propagation by steepening shown in subsection \ref{chap3:fast_mode_propagation}. Moreover, small-wavelength, high-frequency waves are generated around discontinuity and shocks, and this test simulation can investigate the robustness for such small scale waves from PIC to MHD. Initial condition is the same as \citet{brio1988}, and as follows:
\begin{align*}
    &\rho_L = \rho_0, \ \rho_R = 0.125\rho_0, \\
    &u_L = u_R = 0, \\
    &v_L = v_R = 0, \\ 
    &w_L = w_R = 0, \\
    &B_{x,L} = B_{x,R} = 0, \\
    &B_{y,L} = B_{y,R} = 0.75 B_0, \\
    &B_{z,L} = B_0, \ B_{z,R} = -B_0, \\ 
    &p_L = p_0, \ p_R = 0.1p_0. 
\end{align*}
We set $m_{\rm i} / m_{\rm e} = 25$, $n_{{\rm i}, 0, L} = n_{{\rm e}, 0, L} = 200 \ \text{ppc}$, $T_{{\rm i}, L}/T_{{\rm e}, L} = 1$ (with the thermal speed of $v_{{\rm the}, L} = 0.28 c$), and $\omega_{pe} / \Omega_{ce} = 2$. The grid-size ratio is adopted as $\Delta_{\text{MHD}} / \Delta_{\text{PIC}} = 10$. The size of the simulation domain is $50 \times 200, 800$ for the PIC domain and $5 \times 100$ for the MHD domain (corresponding to $50 \times 1000$ PIC grids). The ion inertial length is resolved by about 18 PIC grids. Symmetric boundaries are imposed in $x$ direction, and periodic boundaries in $y$ direction.

\begin{figure}[!t]
    \includegraphics[width=\linewidth]{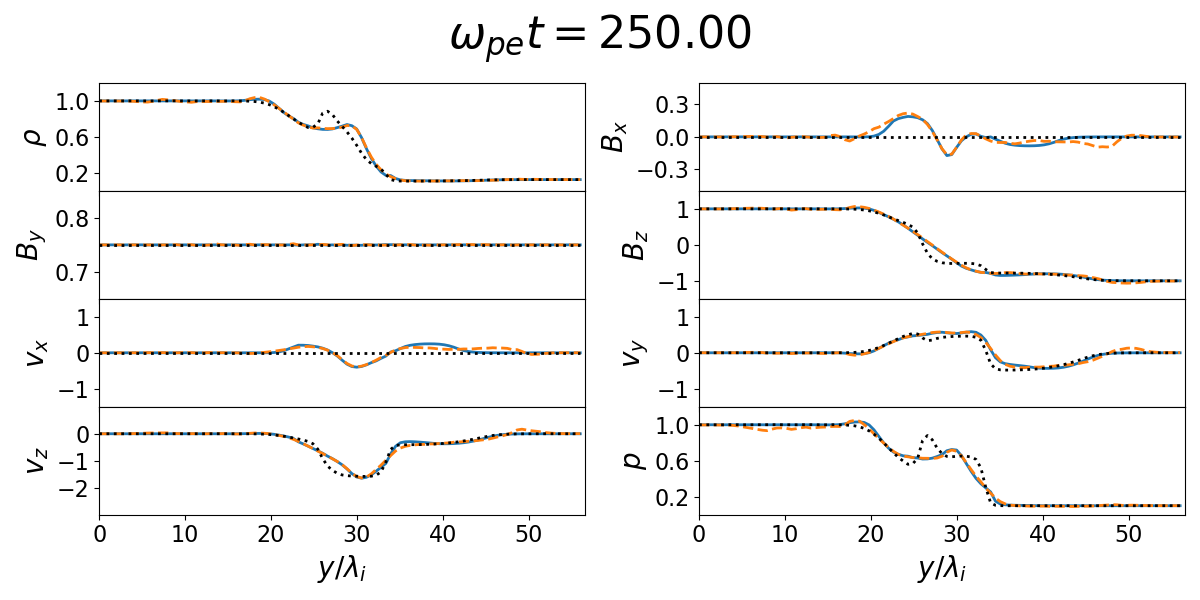}
    \includegraphics[width=\linewidth]{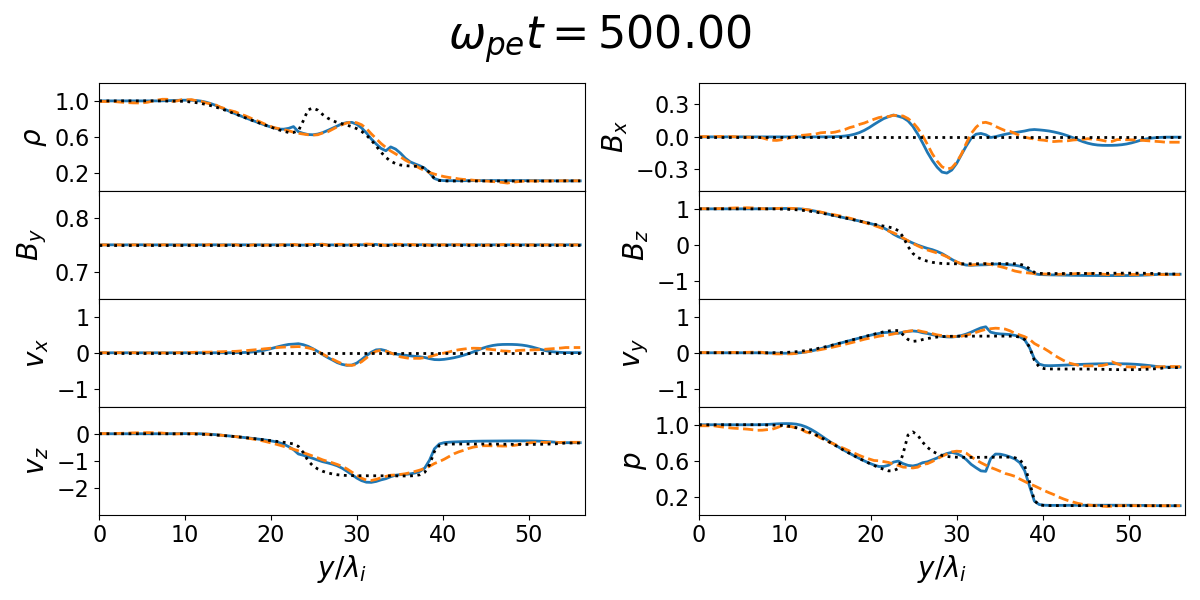}
    \caption{\label{chap3:fig:shock_tube}Snapshots of the shock-tube problem simulations at $\omega_{pe} t = 250$ (top) and $500$ (bottom). The blue line shows the results obtained with 200 PIC grids, while the orange dashed line corresponds to 800 PIC grids. The black dotted line represents the results from the MHD simulation (without the PIC domain). $\rho, B_x, B_y, B_z, V_x, V_y, V_z$, and $p$ are plotted from the top left. 
    {Alt text: The axis shows the position normalized by $\lambda_i$. The range in axis is from 0 to 55. The axis shows the amplitude of each physical quantity.}}
\end{figure}

Figures \ref{chap3:fig:shock_tube} show the snapshots of shock tube problem with the PIC domain sizes of 200 and 800 grids (PIC domain is embedded around $y \sim [23\lambda_i, 34\lambda_i], [6\lambda_i, 51\lambda_i]$), respectively. The results from only MHD simulation are also presented. At $ \omega_{pe} t = 250.00$, rarefaction waves propagate both in the negative ($y \sim 20\lambda_i$) and positive ($y \sim 45\lambda_i$) directions. In both simulations, the structures of rarefaction waves are almost the same as the outcome of MHD simulations (black dotted line). At $\omega_{pe} t = 500.00$ within the region $y = [30\lambda_i, 40\lambda_i]$, the slow shock are clearly seen in $N_{y, \text{PIC}} = 200$ results ($y \sim 40\lambda_i$ is outside of the PIC domain). However, the results for $N_{y, \text{PIC}} = 800$, it appears to show a smoothed structure. Whether slow shocks can stably exist in PIC simulation remains unclear \citep{hau2016, walia2022}. Moreover, in both simulations, the compound wave typically forming near the center in MHD simulation is not observed. Except for slow shock, there is no significant difference between the both simulations in terms of structures in the MHD domain. 

\begin{figure}[htbp]
    \includegraphics[width=\linewidth]{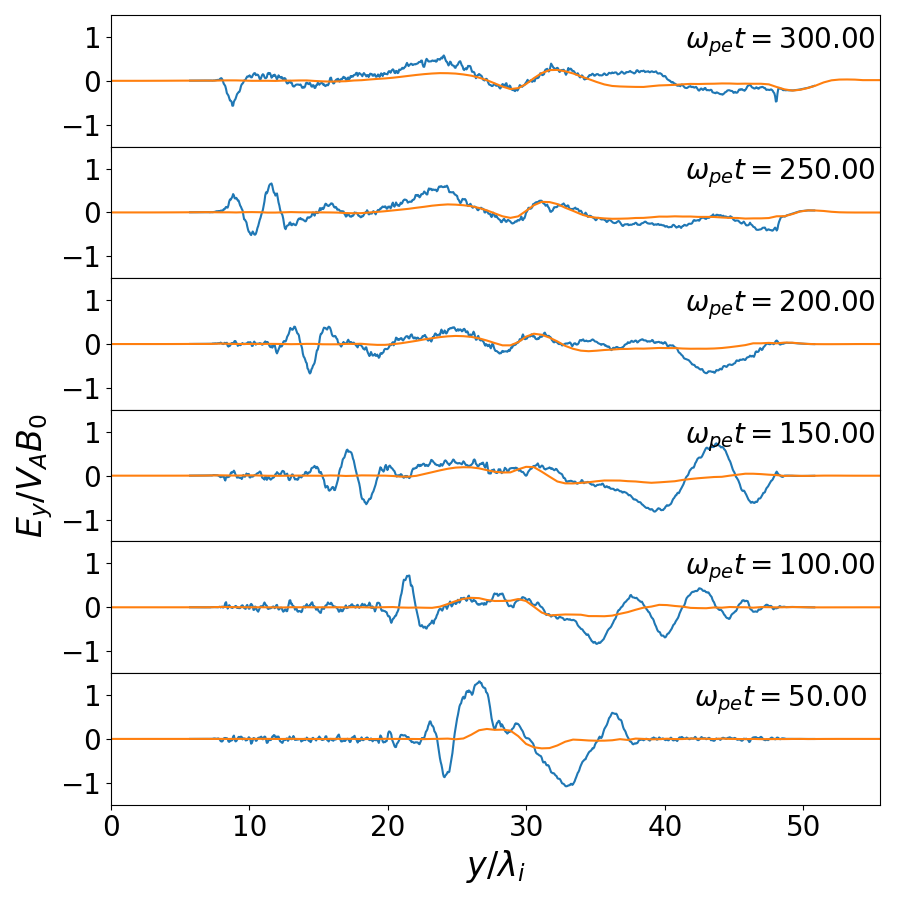}
    \caption{\label{chap3:fig:shock_tube_far_ey}Time evolution of $E_y$ in PIC with a domain size of 800 grids. $E_y$ is normalized by $V_A B_0$ where $V_A:=B_0 / \sqrt{\mu_0 \rho_0}$ is the Alfv\'en speed. Blue line shows the result obtained from PIC, and orange line shows the result obtained from MHD. 
    {Alt text: The axis shows the position normalized by $\lambda_i$. The range in axis is from 0 to 55. The axis shows the amplitude of $E_y$. The range in y axis is from -1.5 to 1.5.}}
\end{figure}

To find out whether small scale waves are existed in the PIC or MHD domain, we checked the electric field which reflect any kinds of waves. Figure \ref{chap3:fig:shock_tube_far_ey} shows the time evolution of $E_y$ from PIC and MHD with a domain size of 800 grids. Short wavelength, high frequency waves are generated around the discontinuity at the center and propagate outward. However, waves corresponding to the oscillations in $E_y$ are not observed in MHD domain. It indicates that the scheme explained in subsection \ref{chap2:multi-hierarchy_scheme} does not transmit small scale waves present in the PIC into MHD. In addition, we find no clear evidence of wave reflection at the interface region. These results indicate that the kinetic waves generated in the PIC domain can penetrate at the interface region, provided that the grid-size ratio is sufficiently large so that small scale waves should not be existed in the MHD domain where the PIC domain embedded.

\section{Application to magnetic reconnection}\label{chap4:magnetic_reconnection}

\begin{figure*}[!t]
    \centering
    \includegraphics[width=0.75\linewidth]{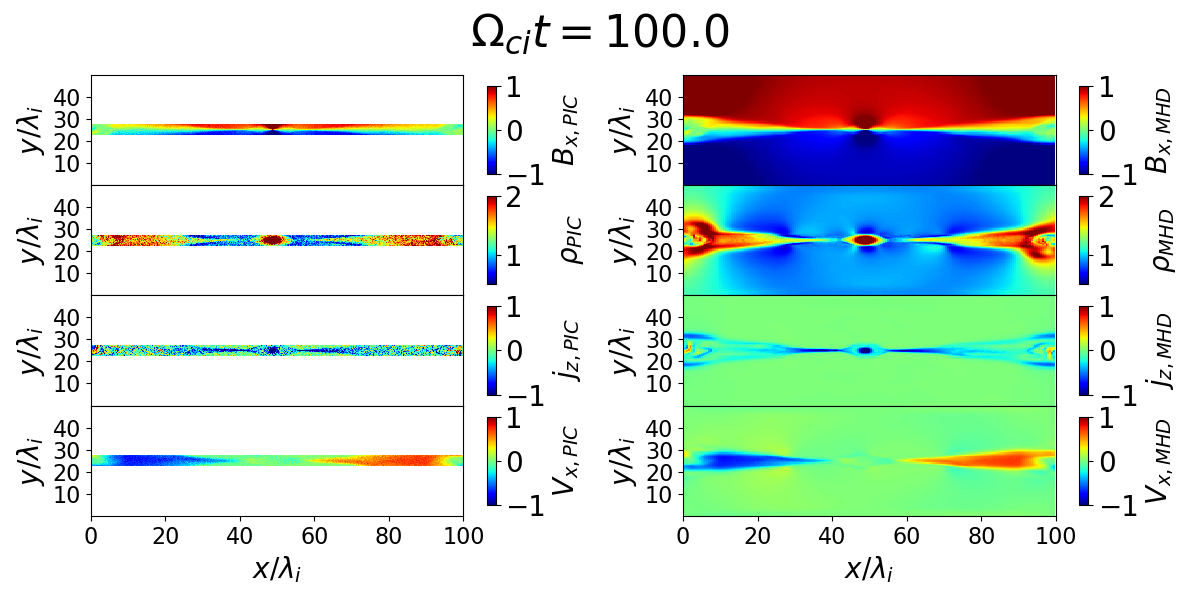}
    \includegraphics[width=0.75\linewidth]{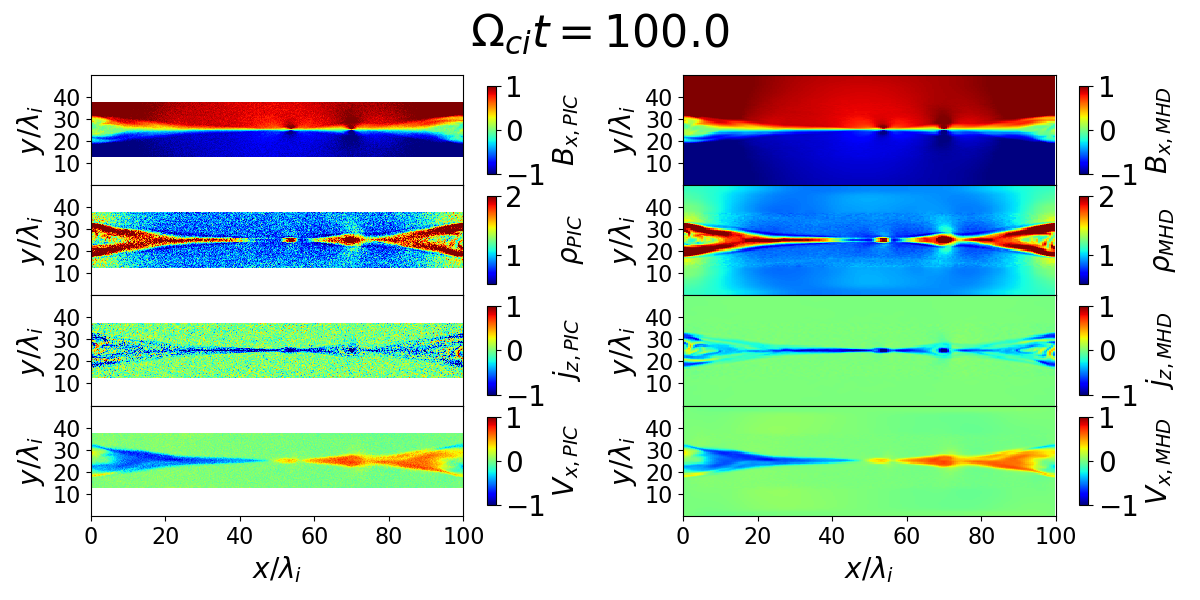}
    \includegraphics[width=0.75\linewidth]{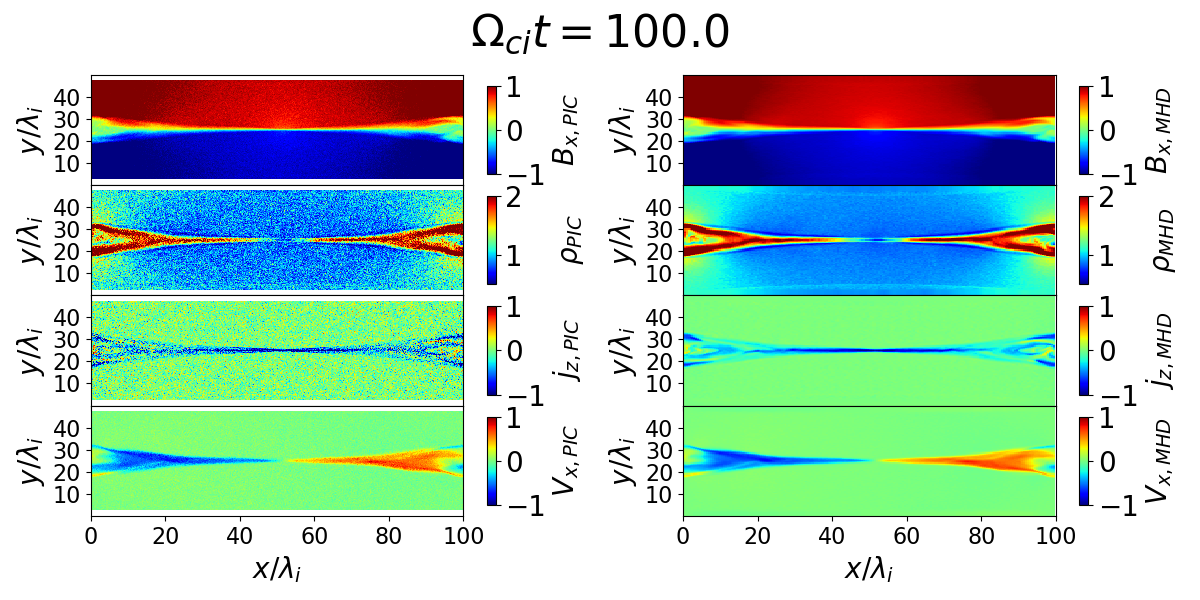}
    \caption{\label{chap4:fig:mrx}Snapshots of magnetic reconnection at $\Omega_{ci} t = 100.0$ for PIC domains of 100, 500, and 900 grids. The upper panel shows the results with the PIC domain size of 100 grids, the middle panel shows those of 500 grids, and the lower panel shows those of 900 grids. The left side shows the results from PIC and the right from MHD. $B_x, \rho, j_z$, and $V_x$ are plotted from the top to bottom. The $B_x$ is normalized by $B_0$, the $\rho$ is normalized by $\rho_0$, the $j_z$ is normalized by $B_0 / \mu_0 \delta$, and the $V_x$ is normalized by $V_A$ where $V_A := B_0 / \sqrt{\mu_0 \rho_0}$ is the Alfv\'en velocity. 
    {Alt text: The $x$ and $y$ axes show the position of $x$ and $y$ normalized by $\lambda_i$. The range in $x$ axis is from 0 to 100, and $y$ axis is from 0 to 50. The contour of the $B_x$ is plotted in the range from -1.0 to 1.0, the $\rho$ is from 0.5 to 2.0, the $j_z$ is from -1.0 to 1.0, and the $V_x$ is from -1.0 to 1.0.}}
\end{figure*}

Taking into account the applicability and limitations of the code revealed by the test calculations in the previous section, we now discuss the application of our multi-hierarchy code to magnetic reconnection. In particular, by varying the size of the PIC domain, we aim to assess the applicability of the code and to clarify whether Hall effect is locally or globally important for the reconnection rate. The initial condition is given by the following force-free current sheet:
\begin{align*}
    &\rho = \rho_0, \\
    &u= 0, \\
    &v= 0, \\ 
    &w= 0, \\
    &B_x = B_0 \tanh(y / \delta),\\
    &B_y = 0,\\
    &B_z = B_0 / \cosh(y / \delta),\\ 
    &p = p_0. 
\end{align*}
We set $m_{\rm i} / m_{\rm e} = 25$, $n_{{\rm i}, 0} = n_{{\rm e}, 0} = 20 \ \text{ppc}$, $T_{{\rm i}, 0}/T_{{\rm e}, 0} = 1$, and $\omega_{pe} / \Omega_{ce} = 1$. The plasma beta is set to 0.25. Grid-size ratio is adopted as $\Delta_{\text{MHD}} / \Delta_{\text{PIC}} = 10$. The size of the simulation domain is $2000 \times 100, 500, 900$ for the PIC region and $200 \times 100$ for the MHD region (corresponding to $2000 \times 1000$ PIC grids). It corresponds to $100\lambda_i \times 50\lambda_i$ simulation box. The current sheet thickness $\delta$ is set to $1\lambda_i$. Reconnection is initiated by a small perturbation in the z-component of the vector potential, given by $\Delta A_z = -2 \delta \exp (-(x^2 + y^2) / (2\delta)^2) \times \epsilon B_0$, where $\epsilon$ is set to 0.1. Periodic boundary condition is imposed in the $x$ direction, while symmetric boundary condition is imposed in the $y$ direction.

The total number of particles in $N_{y, \text{PIC}} = 100$ simulation is decreased to about one-tenth that of $N_{y, \text{PIC}} = 900$. The most computationally intensive part is MPI communication of particles in these simulations, so that the total number of particles determine the total computational time. In conclusion, the total computational time is about one-tenth in $N_{y, \text{PIC}} = 100$ that of $N_{y, \text{PIC}} = 900$.

\begin{figure}[htbp]
    \centering
    \includegraphics[width=\linewidth]{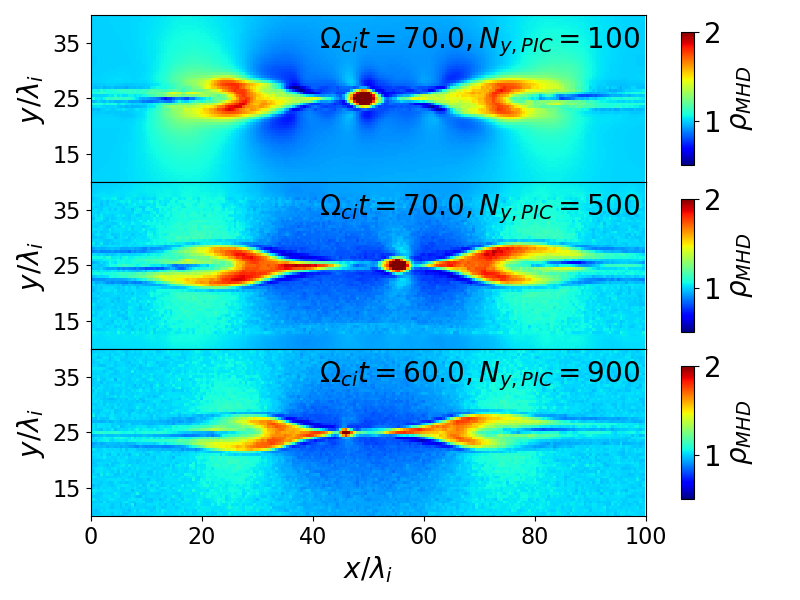}
    \caption{\label{chap4:fig:mrx_rho}Snapshots of the distribution of $\rho$ on the MHD grid. The time of each panel corresponds to the onset of the secondary plasmoid formation: $\Omega_{ci} t = 70.0, 70.0$, and $60.0$ for PIC domain sizes of 100, 500, and 900 grids, respectively. 
    {Alt text: The $x$ and $y$ axes show the position of $x$ and $y$ normalized by $\lambda_i$. The range in $x$ axis is from 0 to 100, and $y$ axis is from 10 to 40. The contour of the $\rho$ is plotted in the range from 0.5 to 2.0.}}
\end{figure}

\begin{figure*}[htbp]
    \centering
    \includegraphics[width=\linewidth]{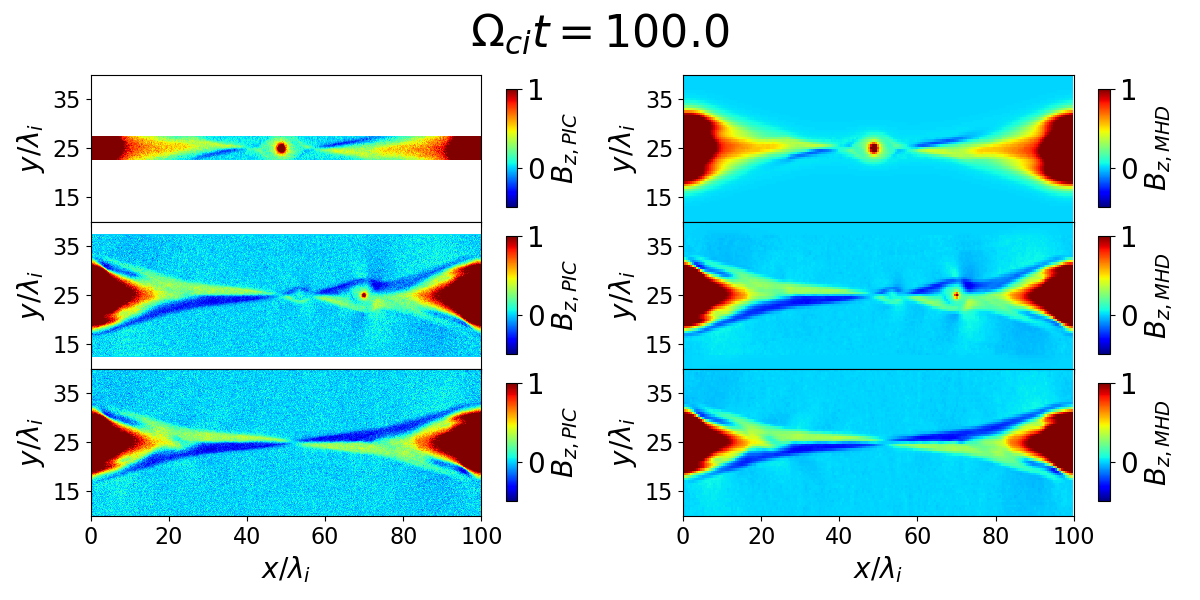}
    \caption{\label{chap4:fig:mrx_bz}Snapshots of the profile of $B_z$ at $\Omega_{ci} t = 100.0$ for PIC domains of 100, 500, and 900 grids. The upper panel shows the results with the PIC domain size of 100 grids, the middle panel shows those of 500 grids, and the lower panel shows those of 900 grids. The left side shows the results from PIC and the right from MHD. $B_z$ is normalized by $B_0$. 
    {Alt text: The $x$ and $y$ axes show the position of $x$ and $y$ normalized by $\lambda_i$. The range in $x$ axis is from 0 to 100, and $y$ axis is from 10 to 40. The contour of the $B_z$ is plotted in the range from -0.5 to 1.0.}}
\end{figure*}

Figure \ref{chap4:fig:mrx} presents the spatial distributions of $B_x$, $\rho$, $j_z$, and $V_x$ at $\Omega_{ci} t = 100.0$ for PIC domains with grid sizes of 100 (top), 500 (middle), and 900 (bottom). The overall structures such as the outflow are similar across the three simulations. In contrast, Figure \ref{chap4:fig:mrx_rho} presents the spatial profile of $\rho$ when secondary plasmoids are formed inside the reconnecting current sheet, appearing at different times. This indicates that the multi-hierarchy simulation of magnetic reconnection is not perfectly identical to the results of the PIC-only simulations. To examine whether this discrepancy solely originates from the PIC domain size, we performed $N_{y,\rm PIC}=100$ simulations with different initial positions and velocities of particles in the PIC domain. The results show that the timing of secondary plasmoid formation differs among these runs. This suggests that the difference of the timing of secondary plasmoid formation may arise not only from a numerical artifact associated with the PIC domain size but also from the random initialization of particles. Moreover, the density $\rho$ profile exhibits a sharp boundary at the MHD–PIC interface, that is the position of sharp structure is different from each simulation. This is evident at $y\sim10, 30\lambda_i$ for the $N_{y,\text{PIC}}=500$ case and at $y\sim5, 45\lambda_i$ for the $N_{y,\text{PIC}}=900$ case, which trace the interface. Such signatures are barely visible in $B_x, j_z$, and $V_x$. The fluctuation of the number density of particles should create these numerical structures.

Figure \ref{chap4:fig:mrx_bz} shows the out-of-plane magnetic field $B_z$ from the three simulations. The quadrupole structure of $B_z$ observed in the reconnection region is caused by the Hall effect, which is a characteristic of collisionless magnetic reconnection \citep{fujimoto2006, drake2009, zenitani2011pop, walia2022}, and cannot be seen in MHD simulation without the Hall term. For $N_{y, \text{PIC}} = 900$, since the PIC region covers almost the whole computational domain, the result closely resembles that of PIC simulation, showing a quadrupole structure in $B_z$. In the simulation with $N_{y, \text{PIC}} = 100$, unlike the cases with $N_{y, \text{PIC}} = 500, 900$, the quadrupole structure gradually disappears outside the PIC region around $y \sim 20, 30\lambda_i$. This is likely because structures generated by kinetic effects (including the Hall effect) do not propagate into the MHD domain, as discussed in the previous section (dispersive kinetic waves such as whistler wave do not propagate into the MHD domain in the test simulations of the Alfv\'en wave propagation).

\begin{figure}[!t]
    \includegraphics[width=\linewidth]{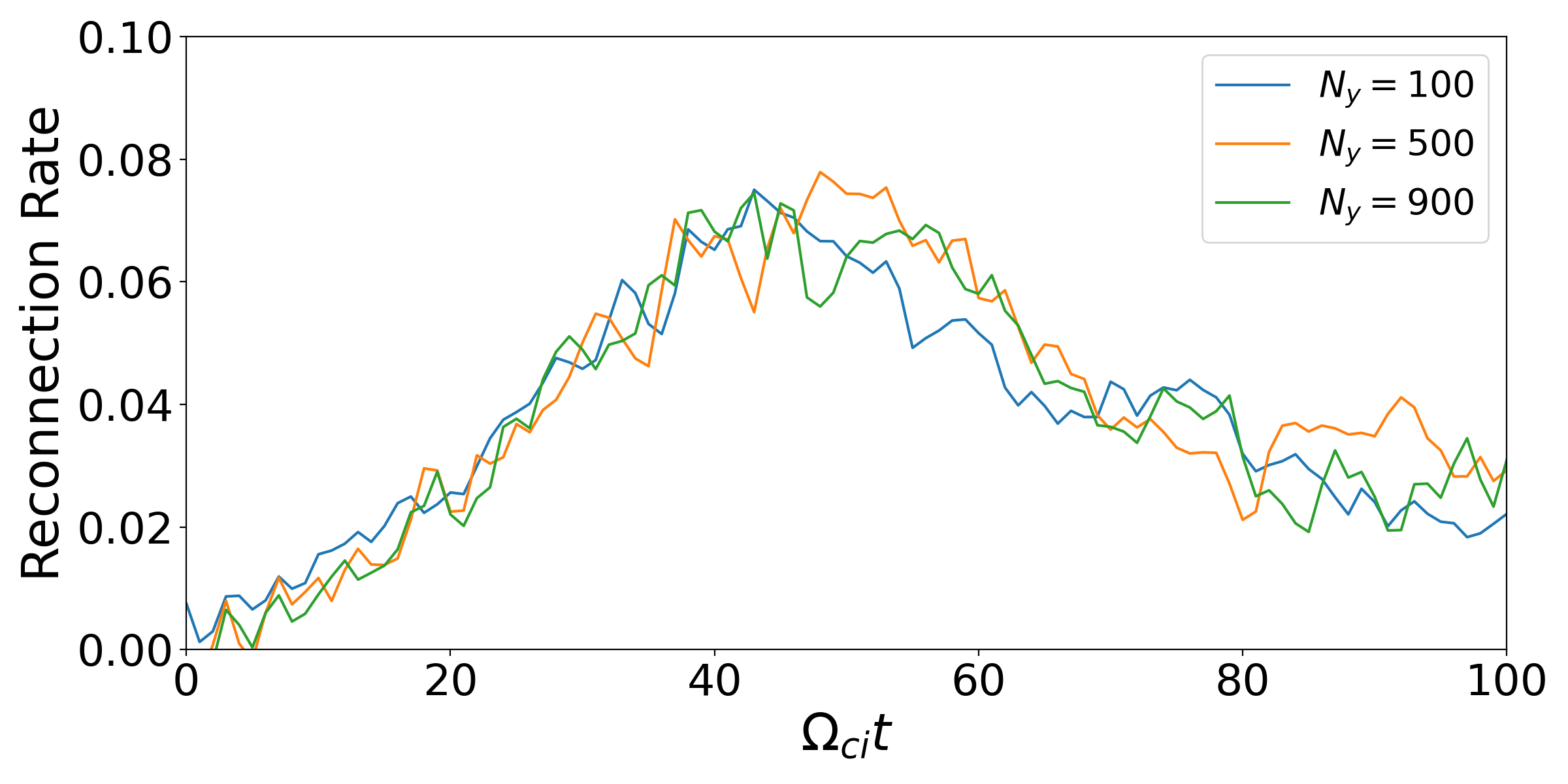}
    \caption{\label{chap4:fig:reconnection_rate}Time evolution of the reconnection rate calculated by Equation \eqref{chap4:eq:reconnection_rate} for PIC domains of 100 (blue), 500 (orange), and 900 grids (green). 
    {Alt text: The $x$ axis shows the time $\Omega_{ci} t$. The range in $x$ axis is from 0 to 100. The $y$ axis shows the value of the reconnection rate. The range in $y$ axis is from 0.00 to 0.10.}}
\end{figure}

For a more quantitative comparison of the three simulations, we evaluate the time evolution of the reconnection rate. The reconnection region is identified as the location of the minimum reconnecting magnetic flux, $\Phi_{\rm min}(t) := \min (\Phi(x, t))$, where $\Phi(x, t) := \int |B_x(x, y, t)| dy$. The reconnection rate is then obtained by 
\begin{eqnarray}
    \text{reconnection rate} = -\frac{1}{V_{A0} B_{0}} \frac{d}{dt}\Phi_{\rm min},
\label{chap4:eq:reconnection_rate}
\end{eqnarray}
where $V_{A0} := B_0 / \sqrt{\mu_0 n_0 (m_i + m_e)}$ is the Alfv\'en speed, and $B_0$ is the anti-parallel magnetic field strength. Figure \ref{chap4:fig:reconnection_rate} shows the time evolution of the reconnection rate from three simulations. We find no significant differences among the three simulation results. The value of the maximum reconnection rate of $\sim 0.08$ is consistent with a previous study using the force-free current sheet \citep{zhou2015}. The results of the $B_z$ structure shown in Figure \ref{chap4:fig:mrx_bz} and the reconnection rate shown in Figure \ref{chap4:fig:reconnection_rate} indicate that the Hall effect seen in $100 \lambda_i$ scale does not affect the reconnection rate.


\section{Summary and Discussion}\label{chap5:summary_and_discussion}

We have developed a multi-hierarchy simulation code that executes PIC simulations in localized regions and MHD simulations in extended regions. From test simulations for the verification of the total energy conservation, Alfv\'en wave propagation, fast mode wave propagation, and shock tube problem, we confirmed that the small scale structures from PIC do not propagate into MHD domain and large scale structures from MHD propagate into PIC domain smoothly. Furthermore, the simulation of magnetic reconnection showed that even when most of the domain is replaced by MHD, the reconnection rate remains consistent with those obtained from PIC simulations. Our results suggest two things; the first is that multi-hierarchy simulation can be applicable to many physical problems, and the second is that the reconnection rate is not significantly affected by the Hall magnetic field which extends $\mathcal{O}(100)\lambda_i$ scale. The inside of the ion diffusion region ($\sim \lambda_i$) should be solved using PIC, however whether the outside is solved has little effect on the reconnection rate.

The test simulations demonstrate the effectiveness of KAMMUY code, as listed below: 
\begin{enumerate}
    \item Energy conservation is satisfied within a few \% error through $\mathcal{O}(10^4)$ PIC steps, especially well conserved when low beta condition is adapted. 
    \item Large scale (MHD scale) waves can propagate through PIC region without distorting its waveform. 
    \item Small scale (kinetic scale) waves hardly propagate into the MHD region where PIC domain is not embedded when using large grid-size ratio of MHD and PIC.
    \item Numerical noises from PIC do not transmit to MHD domain so that multi-hierarchy simulation becomes stable.
\end{enumerate}
An essential aspect of our multi-hierarchy scheme is that the grid size in the MHD domain is much larger than in the PIC domain. As a result, small-scale fluctuations or numerical noise from the PIC domain cannot propagate into the MHD domain. 
On the other hand, the test calculations revealed the following caveats in our code, as listed below: 
\begin{enumerate}
    \item Using small MHD grid size creates unphysical results in MHD domain, for example whistler wave can exist in MHD domain when small $\Delta_{\text{MHD}}$ is used. 
    \item Using small number of particles may create some unphysical error especially for $\rho$ and $p$ which particle noises affects directly.
\end{enumerate} 
Using large $\Delta_{\text{MHD}}$ and number of particles can reduce those limitations. In particular, the test simulation of Alfv\'en wave propagation suggests that the optimal MHD grid size ($\Delta_{\text{MHD}}$) is close to or larger than the ion inertial length $\lambda_i$.

Our multi-hierarchy simulation results of magnetic reconnection indicate that reducing the PIC domain size has only a minor impact on the reconnection rate at least $N_{y, \text{PIC}} = 100$ which corresponds to $5\lambda_i$. It was also found that the structure of the Hall magnetic field depends on the PIC region size, and it was not seen in the MHD region where PIC was not embedded. Although the presence of the Hall term has a significant effect on the magnetic reconnection rate as discussed by \citet{birn2001}, our result suggests that the reconnection rate is not significantly affected by the Hall magnetic field that extends outside the diffusion region in $\mathcal{O}(100)\lambda_i$ systems.
\citet{liu2022nat} presented a theory that the Hall effect inside the ion diffusion region is important for fast magnetic reconnection. The size of the PIC domain in section \ref{chap4:magnetic_reconnection} is at least $5 \lambda_i$, and the ion diffusion region ($\sim \lambda_i$) is inside the PIC domain. It is consistent with the fact that the value of the reconnection rate in multi-hierarchy simulation is almost the same as the previous PIC study \citep{zhou2015}. The fact that the reconnection rate is largely independent of the PIC domain size suggests that our multi-hierarchy method is a promising approach for understanding the physics of the reconnection rate in large systems such as solar flares. Examining whether similar results are obtained for another conditions (e.g., including guide field, 3D systems) when reducing the size of the PIC domain is also our future work. 

Our multi-hierarchy code has the potential to be universally effective in large-scale systems where kinetic effects are locally important. Magnetic reconnection and shocks are typical examples of such processes. In the case of magnetic reconnection, although this study treats only the inflow region with MHD, the outflow region is likewise expected to be handled by MHD. For shocks, all regions except the transition region could potentially be replaced with MHD. In particular, applying MHD to the downstream region would markedly reduce the number of particles, enabling simulations of the long-term shock evolution and making substantial contributions to studies of particle acceleration, especially ion acceleration. Implementing interface region in different direction can further reduce computational cost, which remains as a future work. By incorporating such improvements, the KAMMUY code should facilitate understanding the significance of kinetic effects in large-scale systems such as solar flares.


\begin{ack}
K. A. thanks Dr. Hidetaka Kuniyoshi for deciding our simulation code name KAMMUY from spiritual beings in Ainu culture in Hokkaido, not from the name of a special move in the globally acclaimed Japanese manga NARUTO.
Numerical computations were carried out on GPU cluster at the Center for Computational Astrophysics, National Astronomical Observatory of Japan. It is also partially supported by Recommendation Program for Young Researchers and Woman Researchers, Supercomputing Division, Information Technology Center, The University of Tokyo.
\end{ack}

\section*{Funding}
This work was supported by JSPS KAKENHI Grant Numbers JP24K00688, JP25K00976, JP25K01052, and by the grant of Joint Research by the National Institutes of Natural Sciences (NINS) (NINS program No OML032402).

\section*{Data availability} 
The simulation codes that support the findings of this study are openly available in GitHub repository \url{https://github.com/keita-akutagawa}. The simulation data underlying this article will be shared on reasonable request to the corresponding author.  


\bibliographystyle{aasjournalv7}
\bibliography{bib}

\end{document}